# On the Progenitor of Binary Neutron Star Merger GW170817

LIGO Scientific Collaboration and Virgo Collaboration
(See the end matter for the full list of authors.)

## Abstract

On 2017 August 17 the merger of two compact objects with masses consistent with two neutron stars was discovered through gravitational-wave (GW170817), gamma-ray (GRB 170817A), and optical (SSS17a/AT 2017gfo) observations. The optical source was associated with the early-type galaxy NGC 4993 at a distance of just ∼40 Mpc, consistent with the gravitational-wave measurement, and the merger was localized to be at a projected distance of ∼2 kpc away from the galaxy's center. We use this minimal set of facts and the mass posteriors of the two neutron stars to derive the first constraints on the progenitor of GW170817 at the time of the second supernova (SN). We generate simulated progenitor populations and follow the three-dimensional kinematic evolution from binary neutron star (BNS) birth to the merger time, accounting for pre-SN galactic motion, for considerably different input distributions of the progenitor mass, pre-SN semimajor axis, and SN-kick velocity. Though not considerably tight, we find these constraints to be comparable to those for Galactic BNS progenitors. The derived constraints are very strongly influenced by the requirement of keeping the binary bound after the second SN and having the merger occur relatively close to the center of the galaxy. These constraints are insensitive to the galaxy's star formation history, provided the stellar populations are older than 1 Gyr.

*Key words:* binaries: general – gravitational waves – stars: kinematics and dynamics – stars: neutron

## 1. Introduction

The era of observational gravitational-wave (GW) astronomy was firmly marked by the detection of the first binary black hole coalescence GW150914 (Abbott et al. 2016) by the Advanced LIGO detectors (Aasi et al. 2015). Discovery of a GW source accompanied by coincident electromagnetic (EM) emission, however, remained elusive until now.

On 2017 August 17 the Advanced LIGO (Aasi et al. 2015) and Advanced Virgo (Acernese et al. 2015) interferometer network recorded a transient GW signal consistent with the coalescence of a binary neutron star (BNS) GW170817 (Abbott et al. 2017b). Independently, a gamma-ray signal, classified as a short gamma-ray burst (sGRB), GRB 170817A, coincident in time and sky location with GW170817 was detected by the *Fermi*-GBM instrument (Abbott et al. 2017a, 2017b). The three-detector GW data analysis led to the smallest sky-localization area ever achieved for a GW source: $\simeq$31 deg$^2$ when initially shared with the astronomy LIGO–Virgo partners (LIGO Scientific Collaboration & Virgo Collaboration 2017) and later improved to $\simeq$28 deg$^2$ with a fully coherent data analysis (Abbott et al. 2017b).

Aided by the tight localization constraints of the three-detector network and the proximity of the GW source, multiple independent surveys across the EM spectrum were launched in search of a counterpart beyond the sGRB (Abbott et al. 2017c). Such a counterpart, SSS17a (later IAU-designated AT 2017gfo), was first discovered in the optical less than 11 hours after merger, associated with the galaxy NGC 4993 (Coulter et al. 2017a, 2017b), a nearby early-type E/S0 galaxy (Lauberts 1982). Five other teams made independent detections of the same optical transient and host galaxy all within about one hour and reported their results within about five hours of one another (Allam et al. 2017; Arcavi et al. 2017a, 2017b; Lipunov 2017b; Tanvir & Levan 2017; Yang et al. 2017; Soares-Santos et al. 2017; Lipunov et al. 2017a). The same source was followed up and consistently localized at other wavelengths (e.g., Corsi et al. 2017; Deller et al. 2017a, 2017b, 2017c; Goldstein et al. 2017; Haggard et al. 2017a, 2017b; Mooley et al. 2017; Savchenko et al. 2017; Alexander et al. 2017; Haggard et al. 2017c; Goldstein et al. 2017; Savchenko et al. 2017). The source was reported to be offset from the center of the galaxy by a projected distance of about 10″ (e.g., Coulter et al. 2017a, 2017b; Haggard et al. 2017a, 2017b; Kasliwal et al. 2017; Yang et al. 2017; Yu et al. 2017). NGC 4993 has a Tully–Fisher distance of ∼40 Mpc (Freedman et al. 2001; NASA/IPAC Extragalactic Database[164]), which is consistent with the luminosity distance measurement from gravitational waves ($40^{+8}_{-14}$ Mpc). Using the Tully–Fisher distance, the ∼10″ offset corresponds to a physical offset of $\simeq$2.0 kpc. This value is consistent with offset measurements of sGRBs in other galaxies, though below the median value of ∼3–4 kpc (Fong et al. 2010; Fong & Berger 2013; Berger 2014).

BNS systems were first revealed with the discovery of PSR B1913+16, the first binary radio pulsar ever detected (Hulse & Taylor 1975). This immediately triggered new ideas for how such close pairs of neutron stars can form in nature (De Loore et al. 1975; Flannery & van den Heuvel 1975; Massevitch et al. 1976; Clark et al. 1979), based on models for the formation of high-mass X-ray binaries (van den Heuvel & Heise 1972; Tutukov & Yungelson 1973) and Wolf–Rayet X-ray binaries, for which strong orbital shrinkage is needed

[164] The NASA/IPAC Extragalactic Database (NED) is operated by the Jet Propulsion Laboratory, California Institute of Technology, under contract with the National Aeronautics and Space Administration.

(van den Heuvel & De Loore 1973). With years of pulsar-timing observations PSR B1913+16 provided the first firm evidence that GWs existed (Einstein 1916, 1918) and were emitted by close binary compact objects (Taylor & Weisberg 1982). This discovery greatly motivated the efforts to directly detect GWs with laser-interferometric detectors and made BNS coalescence events key targets in GW searches (see Abadie et al. 2010 for an overview).

The formation of close binaries with two neutron stars that will merge within a Hubble time is now understood to require complex evolutionary sequences of massive binaries that involve stable and unstable mass-transfer phases and two core-collapse supernova (SN) explosions through which the binary system survives (for reviews, see, e.g., Kalogera et al. 2007; Postnov & Yungelson 2014; Tauris et al. 2017). In particular, the SN explosions that lead to the formation of neutron stars are expected to develop asymmetries during the collapse, either due to neutrino emission or an anisotropic explosion (e.g., Kusenko and Segrè 1996; Janka et al. 2007; Janka 2013). This anisotropy imparts linear momentum on the stellar remnant, known as an *SN kick* or *natal kick*.

Strong evidence for this process comes from observations of Galactic pulsar proper motions, which indicate some neutron stars are moving substantially faster than the inferred speed of their progenitors and must receive a large SN kick of $\sim$400–500 km s$^{-1}$ at birth (Lyne & Lorimer 1994; Kaspi et al. 1996; Arzoumanian et al. 2002; Chatterjee et al. 2005; Hobbs et al. 2005; Verbunt et al. 2017). However, comprehensive studies of the known BNS systems in the Milky Way have shown that some neutron stars, particularly those in binary systems, might receive smaller kicks than their isolated counterparts (Podsiadlowski et al. 2004; van den Heuvel 2007).

About a decade after the Hulse-Taylor discovery, mergers of two neutron stars were proposed as a potential source of GRBs (Goodman 1986; Paczynski 1986; Eichler et al. 1989; Narayan et al. 1992), especially those of short duration (Kouveliotou et al. 1993). Since the discovery of host galaxies for short GRBs in 2005 (Berger et al. 2005; Fox et al. 2005; Gehrels et al. 2005; Hjorth et al. 2005; Villasenor et al. 2005), substantial evidence had accumulated in support of this hypothesis. For example, many sGRBs have a significant offset relative to the center of their host galaxy (see, e.g., Troja et al. 2008; Fong et al. 2010; Church et al. 2011; Behroozi et al. 2014): this suggests that the progenitors of these sources have migrated from their birth sites to their eventual explosion sites. Specifically, the offset distribution, together with the locations of sGRBs relative to the stellar light of their hosts, are indicative of systemic kicks (see, e.g., Berger 2014). To date, GW170817 is the strongest observational evidence for an extragalactic BNS system and the first GW signal confidently coincident with an sGRB (Abbott et al. 2017a).

In this study, we focus on constraining the immediate progenitor of GW170817 right before the second SN (SN2) that formed the BNS system. We use (i) SN-kick dynamics and kinematic modeling within the host galaxy from SN2 to merger, and (ii) the GW-measured neutron star masses, the identification of the source host galaxy, and its projected distance offset from the galactic center based on the early optical detections (Section 2). We emphasize that we develop this analysis using the very limited knowledge about the galaxy properties available in the literature prior to the announcement of the GW170817 discovery, as at this time we do not have access to the new analysis of galaxy characteristics and star formation history. We present our main results for constraints on the SN kicks, progenitor masses, pre-SN semimajor axes, and galactic radii of SN2 in Section 3, and we explore the sensitivity of our results to all our input assumptions. We find that the constraints are (i) primarily dictated by the requirement that the progenitor remains bound after SN2 and (ii) insensitive to the star formation history of the host galaxy, provided stellar ages are longer than $\simeq$1 Gyr. In Section 4, we use the GW BNS merger rate to estimate a BNS formation efficiency for NGC 4993, comment on the role of NGC 4993's globular cluster content in BNS formation, and conclude our analysis.

## 2. Analysis Methodology

To investigate the constraints that can be placed on the progenitor of GW170817, we develop a modeling approach comprised of the following elements: (i) assume a gravitational-potential model for the host galaxy, described by a stellar and dark-matter (DM) density profile; (ii) place binary systems in the galaxy according to the stellar profile, and give them a pre-SN orbit in the galaxy; (iii) sample the pre-SN binary properties (pre-SN semimajor axis, progenitor mass of the second neutron star, location of SN2 within the galaxy) and the SN-kick velocity imparted on the binary following from SN2, using multiple assumptions about the underlying distribution of these parameters; (iv) sample the post-SN masses from GW parameter-estimation posterior samples of GW170817; (v) determine if the binary remains bound after SN2 and calculate the post-SN orbital properties, systemic velocity, and inspiral time, assuming two-body orbital mechanics and an instantaneous SN explosion; (vi) evolve the system forward in time, following the trajectory of the binary through the static galactic potential until it merges; (vii) select the systems with a projected offset at merger consistent with the GW170817 measurements, and label them as *GW170817-like*; (viii) impose constraints based on the age at which the binary formed (thus, its delay time between SN2 and merger) and the true (unprojected) distance from the galactic center, and investigate how such constraints affect our inference on progenitor properties; (ix) repeat the above steps for different input assumptions of the progenitor properties to assess the robustness of our results.

For each set of input assumptions, we evolve 50 million binaries according to the above procedures, which is sufficient to properly sample the distributions of GW170817-like systems. This section provides the model details that are adopted in our analysis.

### 2.1. Source Properties

The orbital-dynamics and kinematic analyses presented here require both GW and EM information. The post-SN orbital characteristics of a binary, such as the semimajor axis, eccentricity, and systemic velocity, depend on the component masses of the binary, which are measured in the GW inspiral. The projected offset of the binary relative to NGC 4993's center, measured by EM observations, allows us to select GW170817-like systems in the model populations.

The best-measured property of a GW inspiral is a combination of the component masses known as the *chirp*

*mass*, as it determines the leading-order frequency evolution of a GW signal (Cutler & Flanagan 1994; Blanchet et al. 1995). As the binary orbit shrinks and the orbital period decreases, the GW phase becomes progressively influenced by relativistic effects that are related to the mass ratio. Due to its higher-order contribution, the mass ratio is constrained to a lesser degree than the chirp mass. The measurement of these two parameters are used to extract the component masses of the binary. GW170817 had a measured primary mass of 1.36–1.60 $M_\odot$ and a secondary mass of 1.17–1.36 $M_\odot$, using low-spin priors isotropic in orientation and with $a < 0.05$, where $a$ is the dimensionless spin parameter (see Abbott et al. 2017b for more details). Such low-spin priors are consistent with measured spins in Galactic BNS systems (Brown et al. 2012). We sample posterior distributions of these component mass measurements for each binary realization and assume that the secondary neutron star is the result of SN2.

The location of the source is measured with optical and X-ray observations to an accuracy of $\lesssim 0\farcs5$ (Coulter et al. 2017a, 2017b; Haggard et al. 2017a, 2017b; Kasliwal et al. 2017; Yang et al. 2017; Yu et al. 2017). We combine the information in these references with the range of distances reported in NED and adopt a projected offset distance of $\simeq 2.0 \pm 0.2$ kpc for our analysis.

### 2.2. Galactic Model for NGC 4993

To approximate the galactic potential of NGC 4993, we employ the Hernquist density profile (Hernquist 1990) for the stellar component and the Navarro–Frenk–White (NFW) density profile (Navarro et al. 1996) for the DM halo. We use the stellar profile for sampling the location of binaries within the galaxy, and both the stellar and DM profile for calculating the pre-SN circular galactic velocity and evolving the post-SN binaries in the combined static potential.

The Hernquist profile has a density distribution given by

$$\rho_\star(r) = \frac{M_\star a_{\rm bulge}}{2\pi r (r + a_{\rm bulge})^3}, \quad (1)$$

where $M_\star$ is the total stellar mass and $a_{\rm bulge}$ is a scale length (Hernquist 1990). This profile satisfies de Vaucouleurs $R^{1/4}$ law, an empirical law for the luminosity as a function of radius for early-type galaxies (de Vaucouleurs 1948). Solving Poisson's equation for the gravitational potential yields

$$\Phi_\star(r) = -\frac{GM_\star}{r + a_{\rm bulge}}. \quad (2)$$

The value for the scale length can be computed numerically in terms of the half-light radius ($R_{\rm eff}$) as $a_{\rm bulge} \approx 0.55 R_{\rm eff}$ (Hernquist 1990).

The NFW profile is one of the most commonly used profiles for representing the density distribution of DM halos:

$$\rho_{\rm DM}(r) = \frac{\rho_0}{\frac{r}{R_s}\left(1 + \frac{r}{R_s}\right)^2}, \quad (3)$$

where $\rho_0$ and the scale radius $R_s$ vary from halo to halo (Navarro et al. 1996). Solving Poisson's equation leads to the gravitational potential:

$$\Phi_{\rm DM}(r) = -\frac{4\pi G\rho_0 R_s^3}{r} \ln\left(1 + \frac{r}{R_s}\right). \quad (4)$$

Given a measurement of the DM halo mass, $M_{\rm DM}$, we assume that $M_{\rm DM} \approx M_{200}$, where $M_{200}$ is the mass of the halo enclosed within radius $R_{200}$ at which the density of the enclosed volume is 200 times the critical density of the universe. To determine the value of the constants, we first find the concentration parameter for this volume, $c$, using the empirical expression from Duffy et al. (2008). The two constants are then calculable: $R_s$ is defined as $R_s = R_{200}/c$ and the density parameter $\rho_0$ is calculated by integrating the mass distribution up to $R_{200}$. Though the gravitational potential energy is dominated by the stellar component at small radii (see Figure 1), we use the combined potential when determining the pre-SN galactic velocity and evolving the binary post-SN: $\Phi_{\rm tot}(r) = \Phi_\star(r) + \Phi_{\rm DM}(r)$.

NGC 4993 has a stellar mass of $(10^{10.454}/h^2)\, M_\odot$ (Lim et al. 2017). This stellar mass is derived using K-band luminosity of the galaxy from the 2MASS Redshift Survey (Huchra et al. 2012), and the relationship between stellar mass and K-band luminosity from the EAGLE simulation (Schaye et al. 2015). For our analysis, we use the median value for the Hubble parameter from Planck Collaboration et al. (2016): $h = 0.679$. The DM halo mass for NGC 4993 is $(10^{12.2}/h)\, M_\odot$ (Lim et al. 2017). In addition to stellar and halo masses, we use measurements of the half-light radius of NGC 4993, $R_{\rm eff}$, which is used in the Hernquist profile. The measured value of $R_{\rm eff}$ for NGC 4993 is provided in galaxy surveys (e.g., Lauberts & Valentijn 1989) and was recently reported as 2.8 kpc (Yu et al. 2017), indicating that the merger occurred at a projected distance of $\sim 0.71\ R_{\rm eff}$ from the NGC 4993 center. With the above information, we construct a simple model for the galactic potential of NGC 4993 to be used in our kinematic modeling.

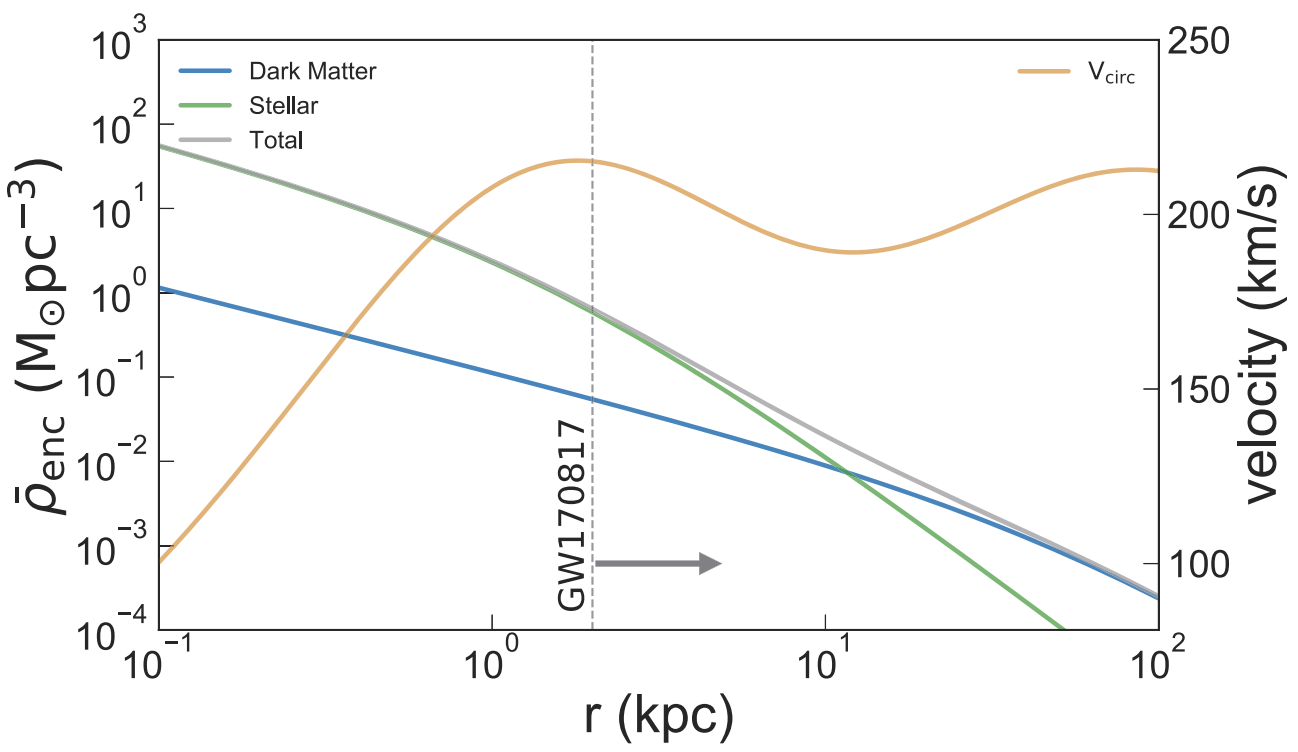

**Figure 1.** Enclosed mass density (left axis, blue/green/gray) and circular velocity (right axis, orange) profiles for our model galaxy. Stellar mass follows a Hernquist profile (Equation (1)) and dark matter an NFW profile (Equation (3)); note that here we plot the average enclosed mass density for a sphere of radius $r$ rather than the mass density at radius $r$. The vertical line marks the projected offset of GW170817, which is a lower limit on the true distance of GW170817 from the center of NGC 4993.

### 2.3. Orbital Dynamics with SN Kicks

We consider the effects of the SN explosion on the orbital dynamics, assuming it is an instantaneous event which imparts a SN kick to the newly formed neutron star and a mass-loss kick (often referred to as a *Blaauw kick*; Blaauw 1961) on the companion neutron star in the binary. We ignore the effects of the first SN (SN1) on the trajectory and orbital properties of the system. The primary reason for this is that previous studies have shown that post-SN1 systemic velocities are small (50–100 km s$^{-1}$) compared to the galactic-motion velocities (see Figure 6 in Belczynski et al. 2002). This is due to the wide pre-SN orbits and hence low pre-SN binary orbital velocities, which regulate the post-SN systemic velocities and limit them to low values (see limits derived in Kalogera 1996). Also, any eccentricity or high orbital separation imparted by SN1 would likely be mitigated by circularization and inspiral during the common-envelope phase of the companion prior to SN2.

The post-SN orbital properties, assuming the binary has circularized prior to SN2, are derived as in Kalogera (1996):

$$A_{\rm post} = G(m_1 + m_2)\left[\frac{2G(m_1+m_2)}{A_{\rm pre}} - V_{\rm kick}^2 - V_{\rm rel}^2 - 2V_{\rm ky}V_{\rm rel}\right]^{-1}, \quad (5)$$

$$1 - e_{\rm post}^2 = \frac{(V_{\rm kz}^2 + V_{\rm ky}^2 + V_{\rm rel}^2 + 2V_{\rm ky}V_{\rm rel})A_{\rm pre}^2}{G(m_1+m_2)A_{\rm post}}, \quad (6)$$

where $A_{\rm pre}$ and $A_{\rm post}$ are the pre-SN and post-SN semimajor axes, $e_{\rm post}$ is the post-SN eccentricity, $m_2$ is the mass of the neutron born in SN2, $m_1$ is the mass of the companion neutron star, $V_{\rm rel}$ is the relative velocity between the binary components pre-SN, and $V_{ki}$ are the components of the SN-kick velocity $V_{\rm kick}$ in the frame of the binary, which is centered on the exploding star with the pre-SN objects lying along the $x$-axis and orbiting in the $x$–$y$ plane.

The system is initially set on a circular orbit in a random direction about the center of the model galaxy. As we show in Figure 2, pre-SN orbits are essential to include when calculating the trajectory of the binary and constraining kick velocities, as kicks tangential to the galactic orbital velocity cause a *slingshot* effect, which is much more efficient at propelling the binary to outer regions of the galaxy than a purely radial kick. In addition, the post-SN systemic velocity of the binary depends heavily on the mass-loss kick as well as the SN kick. Therefore, placing true constraints on the SN kick based on the offset of the merger requires knowledge of the magnitude of this mass-loss kick, which is dependent on the progenitor helium-star mass[165] ($M_{\rm He}$) and pre-SN semimajor axis as well as the final neutron star mass. In Figure 2, we assume an optimally oriented mass-loss kick that is parallel to the galactic velocity to show the true lower limits on the SN kick as a function of SN2 location, for multiple choices of $M_{\rm He}$ and $A_{\rm pre}$. By comparing the solid lines, we see that the lower limit of SN kicks is strongly dependent the progenitor properties we assume. Adopting a fiducial value consistent with our constraints ($M_{\rm He} = 3\,M_\odot$ and $A_{\rm pre} = 2\,R_\odot$) we find that $\gtrsim$99.95% of BNS systems born within 2 kpc of the galactic center satisfy this lower limit. Furthermore, as all systems above this limit reach the offset of 2 kpc in $\lesssim$10 Myr, which is about two orders of magnitude smaller than the typical delay time, it is necessary to continue the evolution of the binary as it explores the galaxy and possibly crosses the projected offset many times, as discussed in Section 2.5.

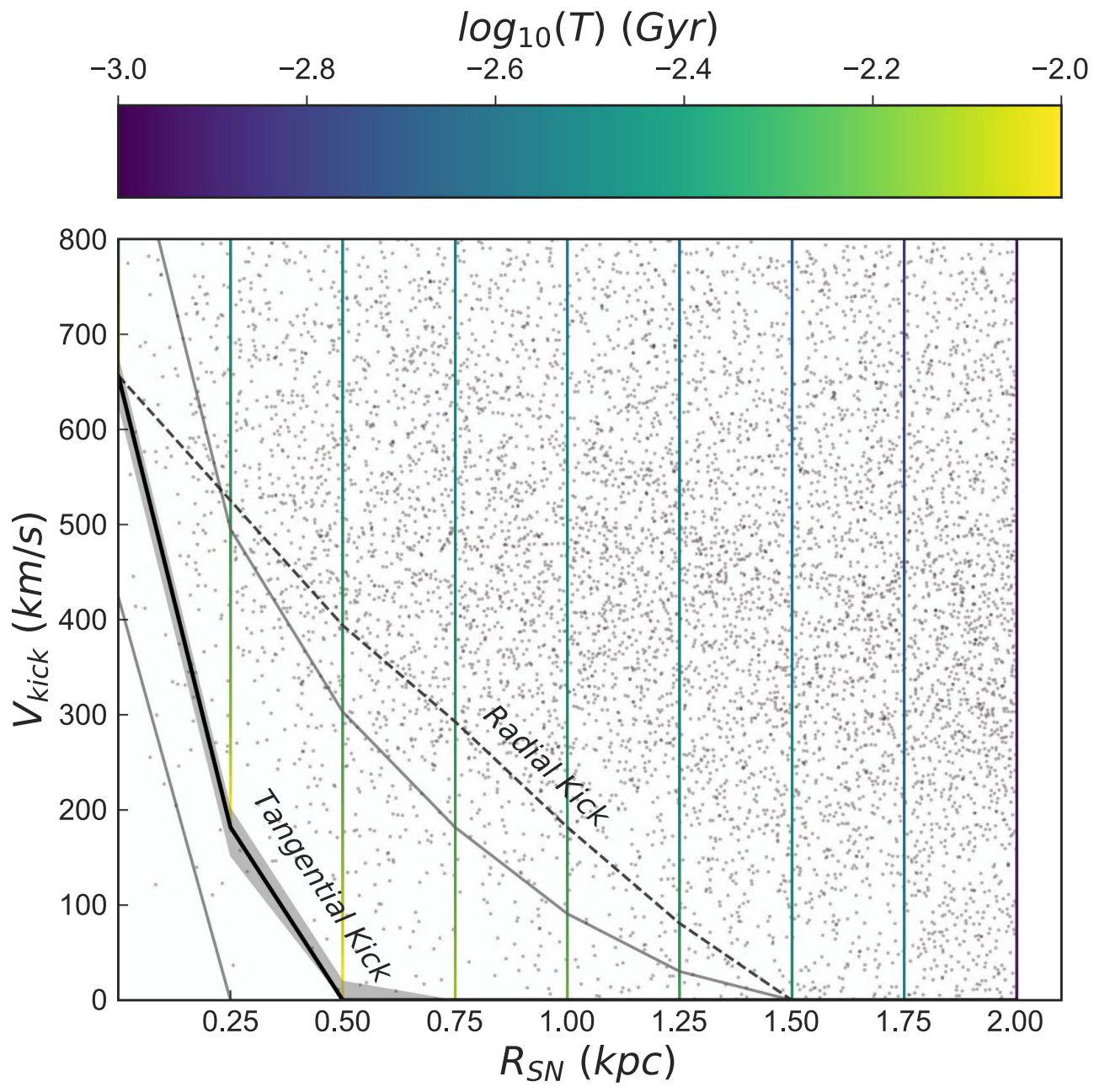

**Figure 2.** Minimum SN-kick velocity required to reach a galactic radius of 2.0 kpc as a function of galactic location at the time of SN2. Mass-loss kicks are accounted for, such that limits can be put solely on the SN kick for a given combination of $A_{\rm pre}$ and $M_{\rm He}$. The thick black line with gray banded region shows the minimum SN kick required to reach 2.0 ± 0.2 kpc when the binary is kicked tangential to the pre-SN galactic velocity, compared to the dashed black line where the binary is kicked radially outward with no contribution from galactic velocity. The assumed fiducial values for this binary progenitor are $A_{\rm pre} = 2\,R_\odot$ and $M_{\rm He} = 3\,M_\odot$. Black points plotted in the background show all sampled systems for various progenitor properties and kick angles as described in Section 2.4; less than 0.05% fall below this limit. The time for systems to reach this offset for various SN-kick velocities is shown by the vertical colored lines. The solid lines to the left and right of the labeled solid line show the tangential SN-kick velocities needed in a more conservative ($A_{\rm pre} = 2\,R_\odot$, $M_{\rm He} = 1.5\,M_\odot$) and less conservative ($A_{\rm pre} = 2\,R_\odot$, $M_{\rm He} = 4.5\,M_\odot$) mass-loss scenario, respectively. These cases all represent a lower limit in the true physical distance that systems must travel to reach a projected distance of 2.0 kpc, as the projected distance from the galactic center is always less than the true distance.

Following the computation of the post-SN orbital properties, the effect of the kick is added to the pre-SN systemic velocity. Due to the SN kick and mass loss, the velocity of the exploding star changes by

$$\boldsymbol{V}_2 = \left(V_{\rm kx},\, V_{\rm ky} + \frac{m_1}{M_{\rm He} + m_1}V_{\rm rel},\, V_{\rm kz}\right), \quad (7)$$

where, again, $M_{\rm He}$ is assumed to leave behind the secondary neutron star component $m_2$. Thus, the contribution of the kicks to the post-SN systemic velocity in the center-of-mass frame of the system becomes

$$\boldsymbol{V}_{\rm sys} = \frac{1}{m_1+m_2}\left(m_2V_{\rm kx},\, m_2V_{\rm ky} - \frac{(M_{\rm He} - m_2)m_1}{m_1 + M_{\rm He}}V_{\rm rel},\, m_2V_{\rm kz}\right). \quad (8)$$

[165] Just before SN2 the companion to the first neutron star is expected to be the He-rich core of a massive star, stripped of its H-rich envelope because of a prior unstable mass-transfer episode and common-envelope phase. Without such a phase, the binary orbits remain too wide for a BNS system that will merge within a Hubble time to form.

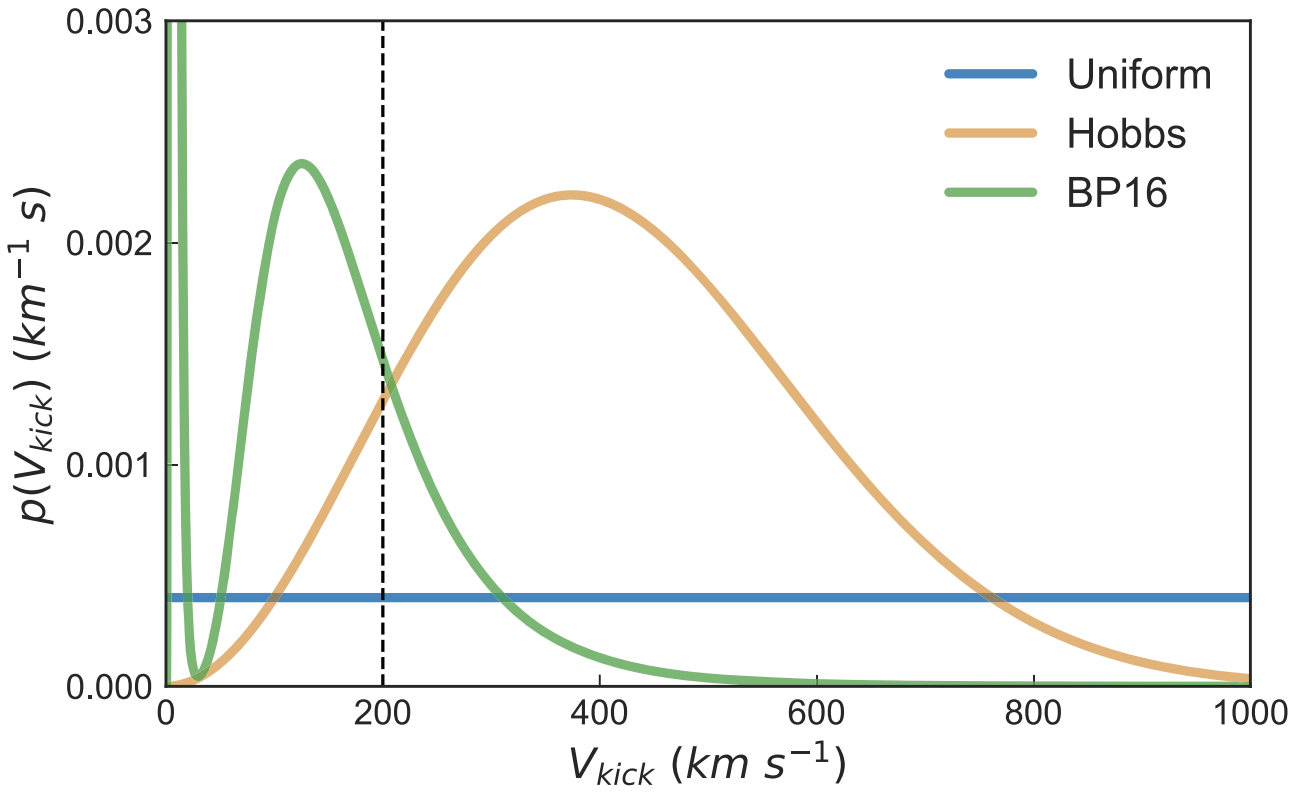

**Figure 3.** Input SN-kick distributions used in this study, as described in Section 2.4. The dashed line indicates a typical galactic orbital velocity in our model of NGC 4993 for comparison; see Figure 1. Note that the distributions are normalized over their full range ([0, 2500 km s$^{-1}$]); we limit the interval that is plotted to better see the morphological differences across distributions.

Given the pre-SN properties of the systems involved, the post-SN systemic velocities are comparable to the galactic-motion velocities (see Figure 1).

Before the systemic velocity is added to the pre-SN galactic velocity at a random angle, we check constraints on the post-SN orbital properties to ensure the system remains bound. First, we require that the post-SN orbit passes through the pre-SN positions of the masses (Flannery & van den Heuvel 1975):

$$(1 - e_{\rm pre}) \leqslant \frac{A_{\rm pre}}{A_{\rm post}} \leqslant (1 + e_{\rm post}). \tag{9}$$

The mass loss and SN-kick magnitude give upper and lower bounds on the amount of orbital expansion or contraction, imposed as in Kalogera & Lorimer (2000):

$$2 - \frac{M_{\rm He} + m_2}{m_1 + m_2}\left(\frac{V_{\rm kick}}{V_{\rm rel}} + 1\right)^2 \leqslant \frac{A_{\rm pre}}{A_{\rm post}} \leqslant 2 - \frac{M_{\rm He} + m_2}{m_1 + m_2}\left(\frac{V_{\rm kick}}{V_{\rm rel}} - 1\right)^2. \tag{10}$$

Finally, the kick velocity is constrained from above by the requirement that the system remains bound, and from below by the minimum kick velocity needed to keep the system intact if more than half the mass of the progenitor is lost in SN2 (Kalogera & Lorimer 2000):

$$\frac{V_{\rm kick}}{V_{\rm rel}} < 1 + \left(2\frac{m_1 + m_2}{M_{\rm He} + m_2}\right)^{1/2}, \tag{11}$$

$$\frac{V_{\rm kick}}{V_{\rm rel}} > 1 - \left(2\frac{m_1 + m_2}{M_{\rm He} + m_2}\right)^{1/2}. \tag{12}$$

### *2.4. Distributions for Pre-SN Progenitor Properties and SN Kicks*

The full 13-dimensional input space from which we sample is

$$\boldsymbol{\Theta} = [m_1, m_2, M_{\rm He}, A_{\rm pre}, R_{\rm gal}, \theta_{\rm gal}, \phi_{\rm gal}, \Omega_{\rm gal}, V_{\rm kick}, \theta_k, \phi_k, \theta_{\rm sys}, \phi_{\rm sys}], \tag{13}$$

**Table 1**
Table of Pertinent Parameters in Our Simulations

| Parameter | Description | Type | Method |
|---|---|---|---|
| $m_1$ | Primary NS Mass | Sampled | GW Parameter Estimation |
| $m_2$ | Secondary NS Mass | Sampled | GW Parameter Estimation |
| $M_{\rm He}$ | Helium-star Mass | Sampled | Uniform, Power Law, *BP16* |
| $A_{\rm pre}$ | Pre-SN Semi-major Axis | Sampled | Uniform, Log Uniform |
| $R_{\rm SN}$ | Galactic Radius of SN2 | Sampled | *Hernquist* |
| $V_{\rm kick}$ | SN Kick Velocity | Sampled | Uniform, *Hobbs*, *BP16* |
| $A_{\rm post}$ | Post-SN semi-major Axis | Calculated | Equation (5) |
| $e_{\rm post}$ | Post-SN Eccentricity | Calculated | Equation (6) |
| $V_{\rm sys}$ | Systemic Velocity | Calculated | Equation (8) |
| $T_{\rm delay}$ | Delay Time | Calculated | Equation (14) |
| $R_{\rm merger}$ | Galactic Radius of Merger | Simulated | N/A |

**Note.** Each parameter is designated as either "Sampled," "Calculated," or "Simulated." *Hernquist* (Hernquist 1990) is a stellar profile used for elliptical galaxies (Section 2.2). *Hobbs* (Hobbs et al. 2005) is a Maxwellian distribution with a scale of 265 km s$^{-1}$ (Section 2.4). *BP16* (Beniamini & Piran 2016) fits log-normal models, with different best-fit parameters for low-eccentricity and high-eccentricity binaries, for distributions in $M_{\rm He}$ and $V_{\rm kick}$ (Section 2.4).

where $m_1$ and $m_2$ are sampled from the posterior parameters of GW170817 (Abbott et al. 2017b); $M_{\rm He}$ is the progenitor helium-star mass; $R_{\rm gal}$, $\theta_{\rm gal}$, and $\phi_{\rm gal}$ are the spherical coordinates of SN2 in the galactic frame of reference drawn from the Hernquist stellar profile; $\Omega_{\rm gal}$ indicates the direction of motion of the system about the center of the galaxy just prior to SN2; $V_{\rm kick}$ is the magnitude of the SN-kick velocity imparted on the newly formed neutron star; $\theta_{\rm k}$ and $\phi_{\rm k}$ are the angular direction of the kick relative to the plane of the binary; and $\theta_{\rm sys}$ and $\phi_{\rm sys}$ are the orientation of the plane of the binary with respect to the galactic coordinates. All angles are sampled isotropically in the sphere. This leaves $M_{\rm He}$, $A_{\rm pre}$, and $V_{\rm kick}$, for which we consider various sampling procedures based on either broad assumptions or observationally motivated distributions.

The majority of constraints on SN kicks come from proper motion measurements of pulsars within our galaxy (Gott et al. 1970; Lyne & Lorimer 1994; Kaspi et al. 1996; Arzoumanian et al. 2002; Chatterjee et al. 2005; Hobbs et al. 2005; Verbunt et al. 2017). We adopt the distribution from Hobbs et al. 2005 (*Hobbs*) as one of our input distributions for $V_{\rm kick}$: a Maxwellian distribution with a 1D rms $\sigma$ of 265 km s$^{-1}$. However, the mechanisms that impart SN kicks to isolated neutron stars may differ from those imparted to neutron stars that remain bound in BNS systems. There are fewer than 20 known BNS systems in the Milky Way, making inference on SN-kick properties a challenging endeavor. Nonetheless, many studies have been performed to better understand the formation process of these systems, combining the observational data with theoretical modeling (e.g., Willems & Kalogera 2004; Piran & Shaviv 2005; Stairs et al. 2006; Willems et al. 2006; Wong et al. 2010; Osłowski et al. 2011; Beniamini & Piran 2016; Tauris et al. 2017). Comprehensive analyses of observed Galactic BNS systems demonstrate that 3–4 systems require small SN kicks ($\lesssim 100\,\mathrm{km\,s^{-1}}$), while another 3–4 clearly require high SN kicks ($\gtrsim 100$–$200\,\mathrm{km\,s^{-1}}$; Wong et al. 2010; Tauris et al. 2017). For the rest, SN-kick constraints are too broad. In addition, theoretical considerations indicate that SN kicks might be smaller for SN2 when progenitors are stripped of their envelopes (Podsiadlowski et al. 2004; van den Heuvel 2007; Janka 2013; Beniamini & Piran 2016). This may suggest bimodality in the SN-kick distribution for neuron stars in binary systems, likely based on the stage of binary evolution the system is in at the time of the SN kick (van den Heuvel 2007).

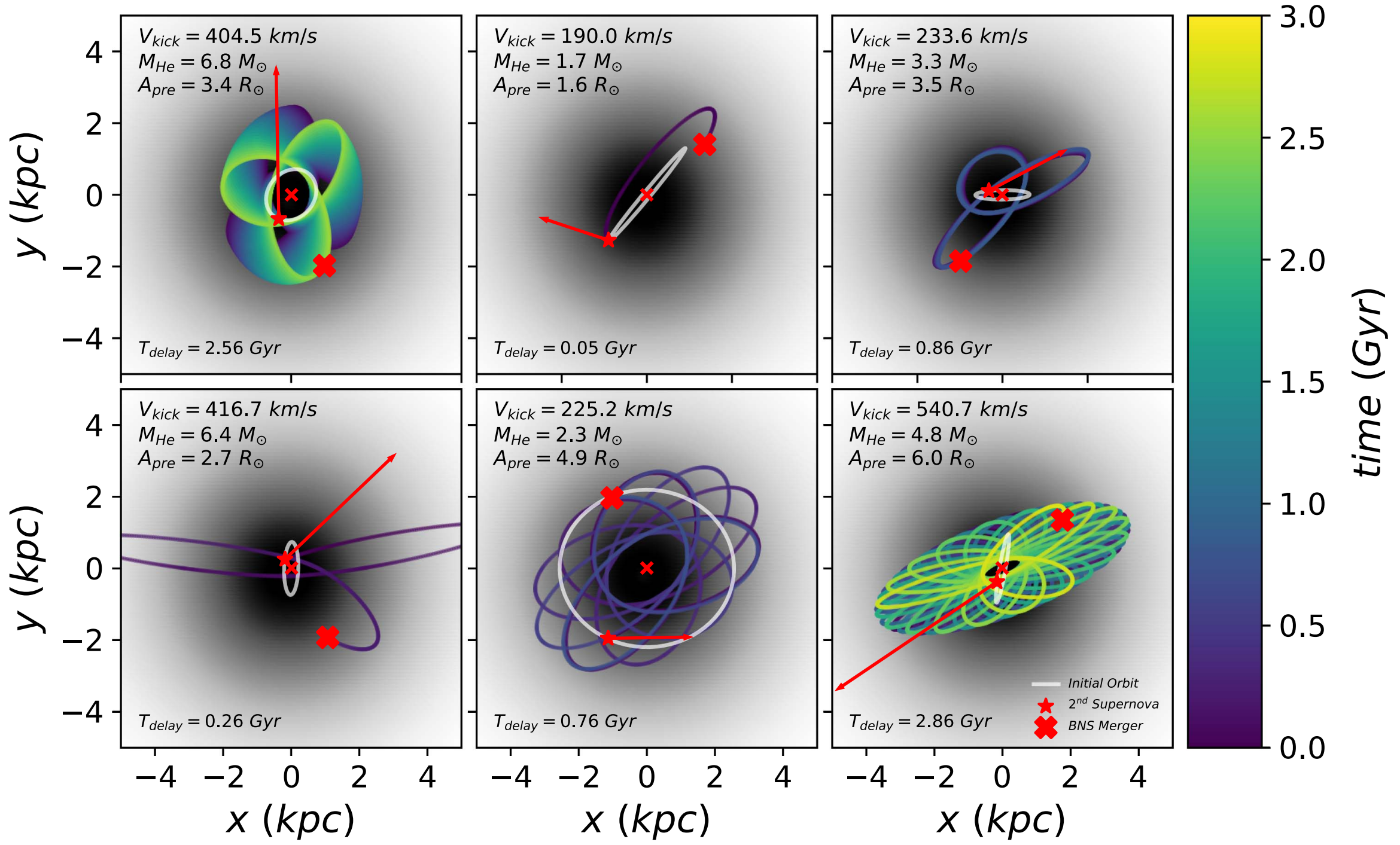

**Figure 4.** Orbital trajectories of representative simulated systems that led to a successful GW170817-like merger. The trajectories show the 2D projection of the orbits that are used to apply the offset constraint from GW170817. The white lines mark the initial (projected) circular orbit of the binary pre-SN, and the red arrows indicate the projected direction of the SN kick. The trajectory of each binary post-SN until merger is displayed on the colored lines, where colors denote the passage of time. Shading follows the projected stellar density of our model galaxy.

Beniamini & Piran (2016) present a two-population model for this apparent bimodality, differentiating low-kick and high-kick Galactic BNSs into two groups based on their observed eccentricity and the rotation period of the pulsar in the system. We use the best-fit parameters from this two-population model (*BP16*) as another kick prescription from which we sample. Beniamini & Piran (2016) also fit a mass-loss model to their two populations, which is tied to the kick model since systems with lower mass loss are expected to have a smaller shell at the time of SN2 and therefore lower SN kicks. We use this two-population model for mass loss as an input distribution for $M_{\rm He}$, which accompanies the bimodal SN-kick model. Physically, the high-kick model corresponds to SN kicks from a Fe core-collapse SN, whereas the low-kick model is meant to emulate the population of binaries that receive electron-capture SN kicks or SN kicks as an ultra-stripped helium star. For the branching ratio between these two populations, we draw 60% of samples from the low-kick model and 40% from the high-kick model, as this is the proportion of Galactic systems that fall into each of these categories (Beniamini & Piran 2016). Finally, we consider an input distribution in SN-kick velocities that is not informed by observations: uniform over the range [0, $2500\,\mathrm{km\,s^{-1}}$] (*uniform*). Figure 3 shows the input distributions of the three SN-kick models described above.

In addition to the various SN-kick velocity input distributions, we consider multiple different sampling procedures for $M_{\rm He}$ and $A_{\rm pre}$. For $M_{\rm He}$, we use a uniform sampling and a power law with an index of $-2.35$ (Salpeter 1955), ranging from $m_2$ (i.e., no mass loss) to the nominal black hole limit of $8\,M_\odot$, along with the two-population maximum-likelihood model for mass loss from Beniamini & Piran (2016). We sample $A_{\rm pre}$ uniform and log uniform from $0.1\,R_\odot$ to $10.0\,R_\odot$. The ranges for both progenitor masses and semimajor axes are motivated by the studies of Galactic BNS systems (e.g., Wong et al. 2010; Tauris et al. 2017).

We summarize the various parameters in our model and sampling procedures in Table 1. To gauge the impact our input distribution on progenitor constraints, we perform runs in which we alter the input distributions of $M_{\rm He}$, $A_{\rm pre}$, and $V_{\rm kick}$ in various ways. We use our least constraining input distribution as our reference: uniform in $V_{\rm kick}$, uniform in $M_{\rm He}$, and uniform in $A_{\rm pre}$. This reference sampling is used for our figures, unless otherwise specified.

### 2.5. *Kinematic Modeling*

With the above we have all necessary quantities to evolve the binary until merger. We calculate the delay time of the binary as a function of post-SN semimajor axis and eccentricity as in Peters (1964):

$$T_{\rm delay}(a_0, e_0) = \frac{15 c^5 k_0^4}{304 G^3 m_1 m_2 (m_1 + m_2)} \times \int_0^{e_0} \frac{e^{29/19}[1 + (121/304)e^2]^{1181/2299}}{(1 - e^2)^{3/2}}\, de, \quad (14)$$

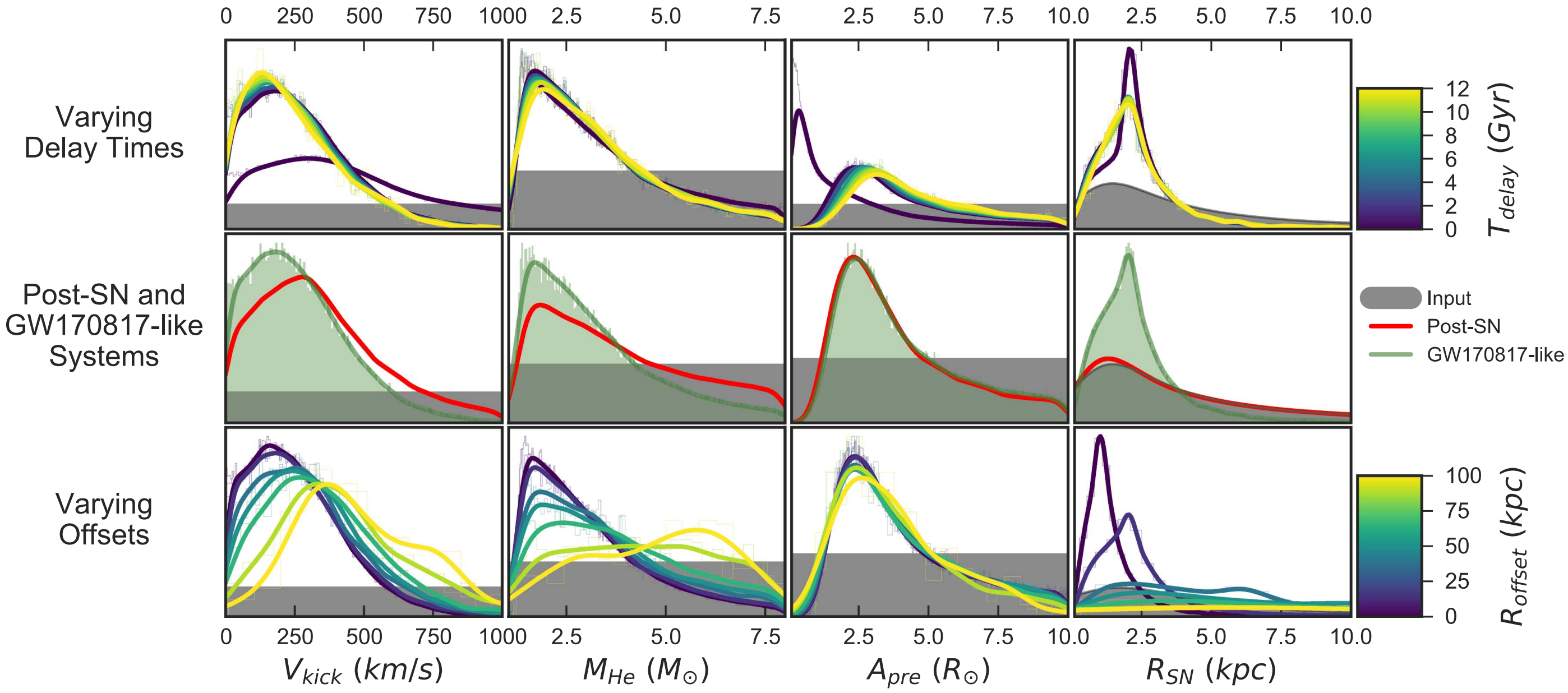

**Figure 5.** Constraints on progenitor properties, SN-kick velocities, and the location of SN2 for various assumed delay times and projected offsets. All plotted lines are kernel density estimates (KDEs) of the recovered distributions, and distributions are normalized over the full range of sampling for a given parameter; vertical axes labels are omitted for readability. In the top row, we set lower limits to the delay times of systems and identify those that match the projected offset of GW170817. As $T_{\mathrm{delay}}$ is coupled to the star formation history of NGC 4993, this has the effect of constraining the simulated stellar population of NGC 4993 to older ages. Sampled distributions are shaded in gray for reference. The middle row shows normalized distributions of binaries that survive SN2 (red) and merge at a projected offset of $2.0 \pm 0.2$ kpc (green; light green shows the histogram of samples to compare with the KDEs). In the bottom row, we investigate how the projected offset of a hypothetical merger similar to GW170817 affects inference on progenitor properties and SN kicks. In the middle and bottom rows, we assume that GW170817 arose from a stellar population older than 1 Gyr.

where $a_0$ and $e_0$ are the initial (post-SN) semimajor axis, and $k_0$ is determined from the initial semimajor axis and eccentricity of the system.

After the binary has evolved for time $T_{\mathrm{delay}}$, we determine the offset of the binary from the center of the galaxy by projecting the system onto the $x$–$y$ plane in galactic coordinates (i.e., we assume the observer is looking at NGC 4993 down the galactic $z$-axis). If the binary ends at an offset between 1.8 and 2.2 kpc and merges in less than a Hubble time, it is considered a GW170817-like system.

We initially take a simplistic approach and assume that all binaries with delay times less than a Hubble time are valid GW170817 analogs. We then consider a full range of possible stellar-population ages for NGC 4993, from as old as the age of the universe to as young as the present. Further discussion on the star formation history of NGC 4993 is found in Section 4. We also vary the projected offset of GW170817, as if it were not known, to investigate how constraints on progenitor properties change as systems are discovered further from their host galaxies.

## 3. Results

Our main results comprise constraints on pre- and post-SN binary properties and SN-kick velocities, which also determine how long each binary lives between SN2 and its GW-driven inspiral and merger. In Figure 4, we show a variety of galactic orbits that potential GW170817 progenitors follow in their host galaxy, depending on post-SN properties and associated delay times. Delay times much longer than the dynamical timescale of the galaxy ($\simeq$20 Myr at 2 kpc) typically lead to progenitors exploring most of the galaxy kinematically despite the merger happening relatively close to the galactic center. Shorter delay times typically lead to simple orbits of minimal structure, facilitating nearby BNS birth and merger locations, although not always (see, for example, the bottom middle panel of Figure 4).

The $T_{\mathrm{delay}}$ times are effectively coupled to the star formation history of NGC 4993, which prior to GW170817 was not well studied. These values are indicative of how long ago SNe typically occurred, and therefore mark the ages of the most dominant stellar populations in this galaxy. In the analysis of our results, we consider a range of different $T_{\mathrm{delay}}$ constraints and assess the sensitivity/robustness of derived constraints on progenitor properties to assumptions about the stellar age of NGC 4993, i.e., $T_{\mathrm{delay}}$ of GW170817-like progenitors. Though the projected offset of the optical counterpart to GW170817 was well constrained, we also consider our results' robustness against this location constraint. Last, we explore different assumed distributions for the initial progenitor properties and SN kick, and we assess the robustness of our results against such changes.

The main results are presented in Figure 5, for our fiducial simulation where we assume uniform distributions for all input parameters (see Section 2.4). For the progenitor populations in the top row, we examine probability density functions (PDFs) on GW170817 progenitor properties when we impose the projected distance offset constraint of $2.0 \pm 0.2$ kpc, and different lower limits on the $T_{\mathrm{delay}}$. It is remarkable that, provided the stellar population in NGC 4993 is older than 1 Gyr, the progenitor constraints are highly robust. We also find this insensitivity to the fine details of stellar ages to be true for our other input distributions and when we constrain $T_{\mathrm{delay}}$ to specific ranges rather than imposing lower limits. Only if $T_{\mathrm{delay}}$ values shorter than 1 Gyr are allowed (i.e., recent star formation has persisted in the host galaxy) are the constraints on the SN kick and the pre-SN semimajor axis strongly affected: shorter time delays imply tighter post-SN BNS systems which allow for tighter pre-SN binaries that can remain bound even with higher SN-kick magnitudes. Delay times shorter than 1 Gyr also produce a very sharp peak in the galactic radius of SN2 ($R_{\rm SN}$) around the merger distance, as the progenitor population becomes dominated by binaries that are born as BNSs relatively close to their merger site with short $T_{\rm delay}$. To summarize, for delay times greater than 1 Gyr, the median values and 90% ranges for our reference sampling are: $\simeq 3.0^{+3.5}_{-1.5}\,M_\odot$ for the progenitor mass of the second neutron star at explosion, $\simeq 3.5^{+5.0}_{-1.5}\,R_\odot$ for the pre-SN semimajor axis, $\simeq 300^{+250}_{-200}$ km s$^{-1}$ for the second SN-kick magnitude, and $\simeq 2.0^{+4.0}_{-1.5}$ kpc for the birth radius away from the galaxy center. More detailed results examining additional parameters and parameter correlations can be found in Figure 8 in the Appendix.

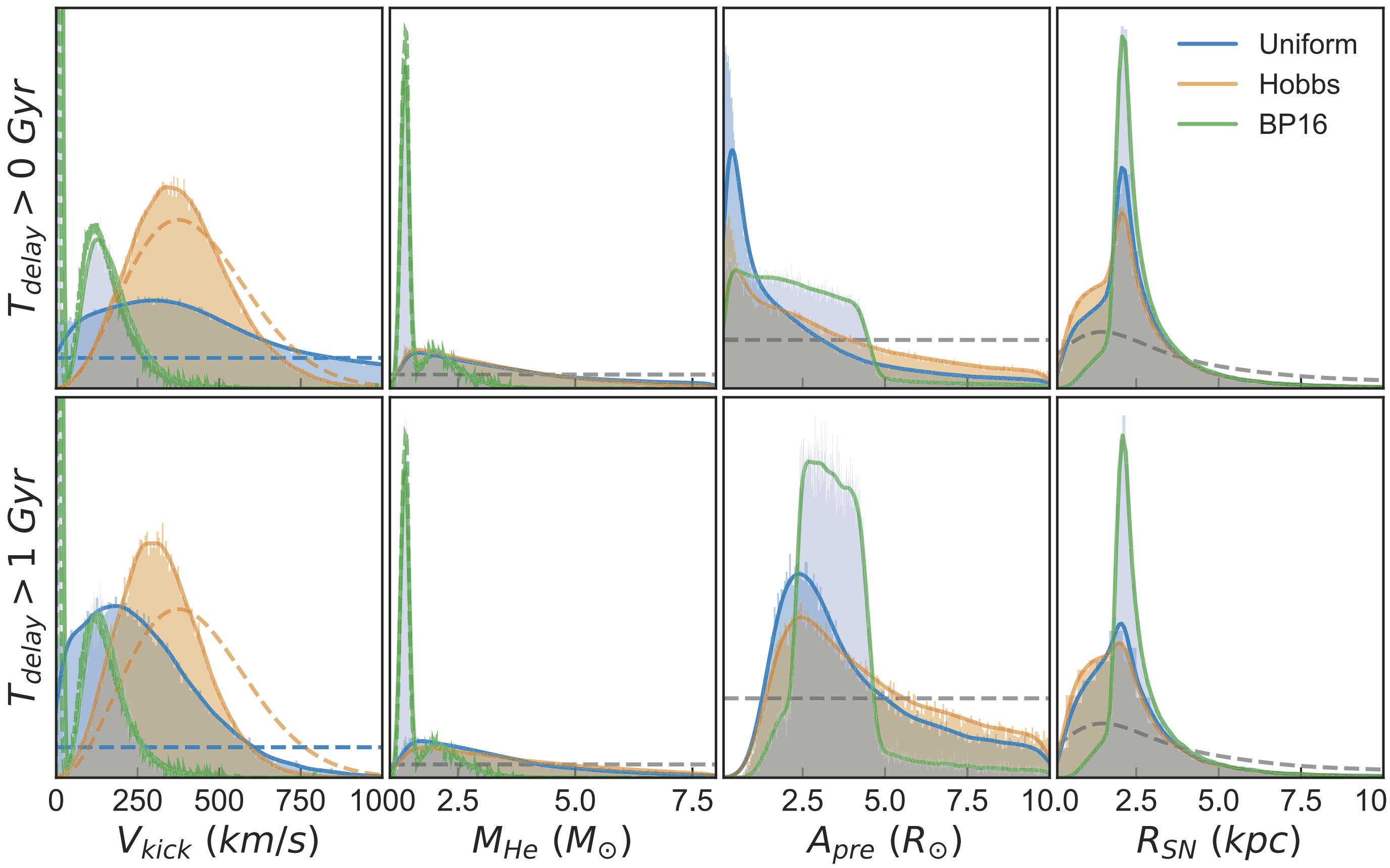

**Figure 6.** Comparison of recovered PDFs from various input distributions on the SN kick. Vertical axes are normalized PDFs for a given parameter and PDFs are normalized over the full range of the parameter; vertical labels are omitted for readability. Blue are the resultant PDFs from kicks drawn uniformly, orange PDFs are kicks drawn from the Hobbs et al. (2005) prescription, and green PDFs are drawn from the two-population Beniamini & Piran (2016) prescription. The top row considers all systems that merge within a Hubble time, and the bottom row systems with delay times bounded by 1 Gyr < $T_{\rm delay}$ < 14 Gyr. Dashed lines show the input distributions for each kick prescription, as well as the input distributions for $M_{\rm He}$, $A_{\rm pre}$, and $R_{\rm SN}$. Note that the input distribution on $M_{\rm He}$ differs for *BP16*, and the input distributions for $A_{\rm pre}$ and $R_{\rm SN}$ are identical across all three models.

In addition to SN-kick velocities, we examine constraints on the post-SN systemic velocities ($V_{\rm sys}$). We find somewhat tighter constraints on $V_{\rm sys}$ for GW170817-like binaries, peaking at $\simeq$250 km s$^{-1}$ and with 90% of systems below $\simeq$400 km s$^{-1}$ when we constrain the population to $T_{\rm delay} \geqslant 1$ Gyr. Tighter constraints are to be expected as the systemic velocities are limited by the requirement that the post-SN binary remains bound; as a result, the systemic velocities saturate at values of about 1.5–2 times the pre-SN relative orbital velocities (see Kalogera 1996 for the analytical derivation of upper limits). We again find that $V_{\rm sys}$ is robust to age constraints; the PDFs on $V_{\rm sys}$ are practically identical provided the stellar population is $\gtrsim$1 Gyr.

In the middle and bottom rows of Figure 5, we examine how significant of a constraint is the knowledge of the merger's offset from the galaxy's center. Results in the middle row demonstrate that the primary origin of our constraints on SN kicks and progenitor properties stems from the requirement that systems remain bound after the explosion. Higher kicks, more massive helium-rich progenitors, and wider pre-SN orbits tend to disrupt a larger fraction of systems. We also find that *any* offset constraint at all differentiates the $R_{\rm SN}$ distributions between SN survivors and GW170817-like systems the most: remaining bound post-SN is not affected by galactic location and without the offset, of course, progenitors follow the galaxy mass distribution. Imposing an offset constraint limits the birth radius to within a factor of typically $\sim$2–3 from the offset. The relatively small offset from the galaxy center shifts the SN kicks and helium-star masses to smaller values, effectively reducing the BNS post-SN systemic velocities, while it leaves the constraints on $A_{\rm pre}$ unaffected.

We further explore the robustness of our results on the assumed input distributions for $V_{\rm kick}$, $M_{\rm He}$, and $A_{\rm pre}$, again adopting the merger projected offset constraint of $2.0 \pm 0.2$ kpc. Specifically in Figure 6, we show our results for the three different SN-kick distributions (see Section 2.4). We choose only two cases of $T_{\rm delay}$ constraints given the robustness of our results demonstrated in Figure 5: (i) no constraint (top row), i.e., star formation has continued in this galaxy up until the present, and (ii) $T_{\rm delay} > 1$ Gyr (bottom row), i.e., the stellar population in NGC 4993 is older than 1 Gyr. It is evident that the constraints on $M_{\rm He}$, $A_{\rm pre}$, and $R_{\rm SN}$ are robust against these different SN-kick assumptions. However, the robustness against different kick assumptions comes with the corollary that the data from this one observation is not extremely informative on the true underlying SN-kick distribution. In general, we see that GW170817 constraints exhibit mild sensitivity to the input SN-kick distributions. The *uniform* and *Hobbs* sampling procedures tend to shift to smaller SN-kick magnitudes but by relatively small amounts. The behavior with the *BP16* assumption is different, but not surprising: the *BP16* input distributions are extremely narrow and prescriptive, strongly dictating the allowed SN kicks and progenitor masses. The constraints for $A_{\rm pre}$ and $R_{\rm SN}$ are slightly stronger than those for $V_{\rm kick}$ and $M_{\rm He}$; $A_{\rm pre}$ is strongly influenced by the limits placed on delay times, and $R_{\rm SN}$ by the offset of GW170817 with respect to the galactic center.

Lastly, we have performed additional simulations with varying assumptions about the neutron star mass posteriors (Section 2.1) and the galaxy parameters (Section 2.2). To test the sensitivity of our results to neutron star mass measurements, we sampled the high-spin prior ($a < 0.89$) component mass posteriors, which have a much broader range of 1.36–2.26 $M_\odot$ and 0.86–1.36 $M_\odot$ for the primary and secondary components masses, respectively (Abbott et al. 2017b). We find quantitatively insignificant differences in our progenitor constraints. We also assess the robustness of our results against variations in the measured properties of the galaxy (i.e., stellar and DM halo masses, effective radius) and find insignificant changes for variations up to $\sim$30%.

We quantitatively test the robustness of our results against all assumption variations by calculating the Kullback–Leibler (KL) divergence (Kullback & Leibler 1951), described in more detail in the Appendix. The KL results, as well as the median and 90% credible intervals on all progenitor parameters and all input distributions are reported in Table 2 in the Appendix, and quantitatively justify our statements on insensitivity and robustness. The median values for the progenitor masses and semimajor axes are mostly consistent with favored values found with forward population synthesis of binary evolution (K. Breivik 2017, private communication).

Prior to SN2, the helium star may have been overflowing its Roche lobe and transferring mass to its neutron star companion. If so, the helium star could have lost significant amounts of mass ($\gtrsim$1 $M_\odot$) prior to its explosion (Pols & Dewi 2002; Ivanova et al. 2003). To investigate the possibility of Roche-lobe overflow (RLO) at the time of the SN2, we examine whether the properties of GW170817 progenitors satisfy the conditions for RLO, adopting the analytical fit for the helium-star radius from Kalogera & Webbink (1998). Figure 7 plots successful binaries on progenitor $A_{\rm pre}/M_{\rm He}$ space, indicating those that would have been in an RLO phase at SN2 in green. We see that a significant fraction ($\simeq$46%, assuming uniform input distributions as described in Section 2.4) of the GW170817 progenitor systems may have been undergoing RLO at the time of the BNS formation. This is not a major surprise, as it is well established by several independent studies that the double pulsar (and other known BNS systems) was also in an RLO phase at the time of SN2 (Willems & Kalogera 2004; Piran & Shaviv 2005; Stairs et al. 2006; Willems et al. 2006; Wong et al. 2010; Beniamini & Piran 2016; Tauris et al. 2017).

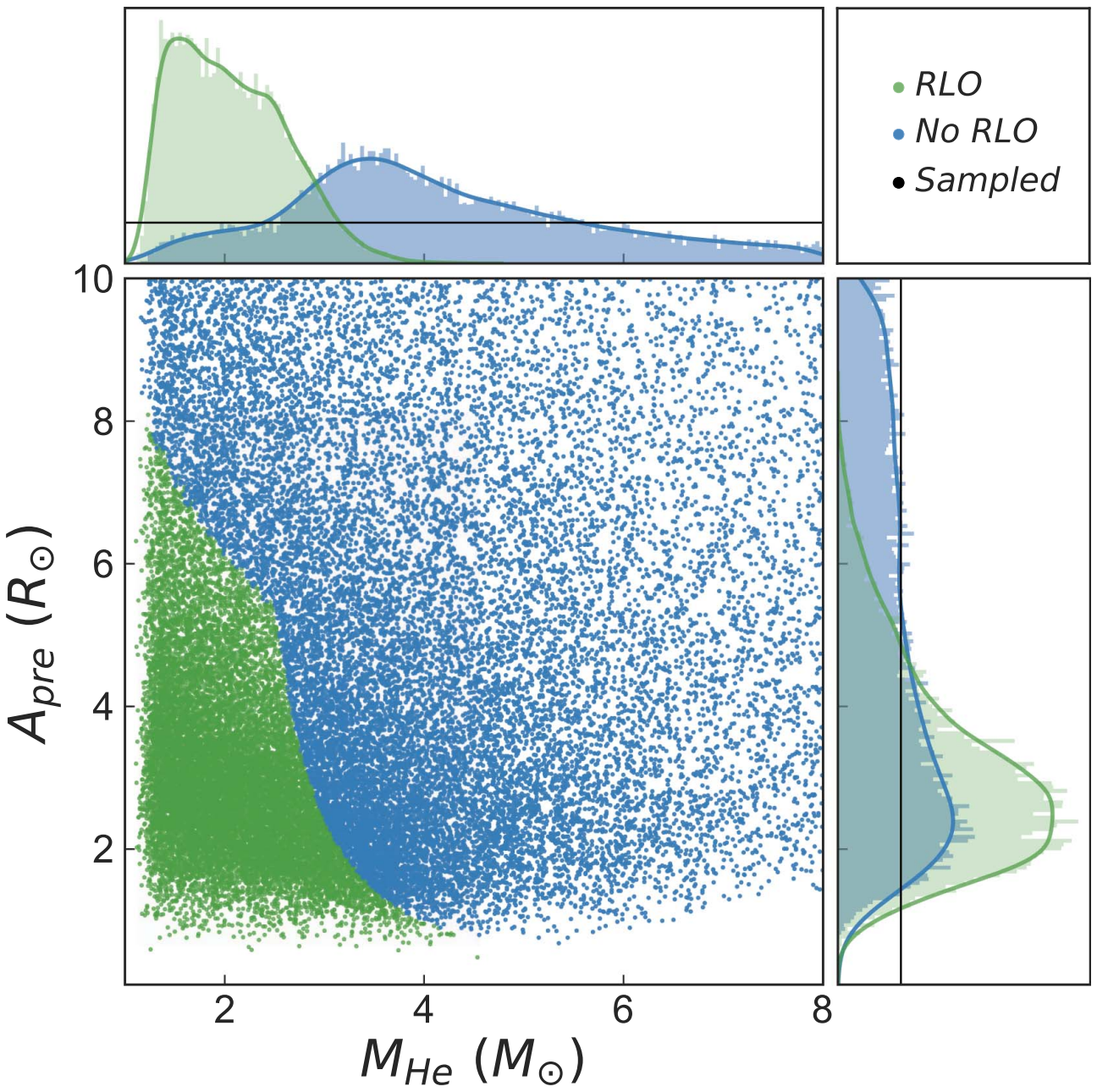

**Figure 7.** Presence of RLO in progenitor systems prior to SN2. Only systems that produced successful analogs of GW170817 are plotted. Green indicates that the system was in RLO prior to SN2, and blue indicates that the system was at a large enough pre-SN semimajor axis to not be experiencing RLO prior to SN2.

## 4. Discussion and Conclusions

In the modeling analysis presented here, we focus on constraining the immediate progenitor of GW170817, from its actual formation at the time of the second SN to the final merger. We use (i) SN-kick dynamics and kinematic modeling within the host galaxy and (ii) the GW-measured neutron star masses, the identification of the source host galaxy, and its projected distance offset from the galactic center based on the early optical discoveries. We make the most minimal/agnostic assumptions possible and avoid full, high-fidelity population synthesis models, which can account for the complex binary evolution before SN2. We explore the robustness of our results for different input assumptions.

In our analysis, we assume that the GW170817 progenitor evolved as an isolated binary in the galaxy's field population. There are no reported results regarding observations of globular clusters (GCs) in NGC 4993, so the number of GCs in the galaxy is not known. Given that NGC 4993 could have a sizable population of GCs, a dynamical formation channel for the coalescing BNSs cannot be ruled out a priori. Typically, the number of observed GCs in a galaxy correlates with the luminosity of the galaxy (Barr et al. 2007; Harris 2016). This observed correlation can be used to estimate the number of GCs in NGC 4993. With an apparent V-band magnitude of 12.4 mag (Bellini et al. 2017) and a distance of 40 Mpc, we find an absolute V-band magnitude of $-20.6$ mag, which for an E/S0-type galaxy would correspond to $250^{+750}_{-150}$ GCs. However, since in general GCs comprise only a small fraction of the total mass of the galaxy ($\sim$0.01%–0.1%; Harris et al. 2015) and estimated observational merger rates for BNSs originating from GCs are low (Grindlay et al. 2006; Ivanova et al. 2008), the isolated formation scenario is preferred. Bae et al. (2014), for example, estimated detection rates for LIGO–Virgo of 0.024 to 0.1 events per year for 1.4 $M_{\odot}$–1.4 $M_{\odot}$ BNSs coming from GCs (assuming a design-sensitivity BNS range of 200 Mpc for the LIGO detectors), which gives $\sim 10^{-4}$–$10^{-2}$ events per year given the sensitivity at the time of detection (∼50, 100, and 25 Mpc for Hanford, Livingston, and Virgo, respectively). Such a low rate estimate is in contrast with the rate implied by the current GW discovery, and therefore we consider it unlikely that GW170817 was formed in a GC.

We can use the current knowledge from just this one BNS detection in GWs to extract a first estimate of the BNS formation efficiency: the fraction of massive binaries that become merging BNS systems. Our GW data analysis has yielded a measurement of the BNS rate density of 320–4740 $\mathrm{Gpc^{-3}\,yr^{-1}}$ (Abbott et al. 2017b), consistent with the measurements from radio-pulsar observations (e.g., Abadie et al. 2010). Given the volume density of Milky Way–like galaxies in the local universe (i.e., galaxies of comparable mass to the Milky Way) this rate measurement translates to 32–474 $\mathrm{Myr^{-1}}$ per Milky Way equivalent galaxy (Abadie et al. 2010). NGC 4993 is of a transitional galaxy type rather than a spiral galaxy. We therefore use the galaxy's stellar mass (instead of their blue luminosities as traditionally done for star-forming galaxies) to scale the BNS rate volume density to a rate for this specific galaxy. For the Milky Way and NGC 4993 the masses are very comparable: both have approximately 60 billion solar masses in stars (Licquia & Newman 2015; Lim et al. 2017), and therefore the BNS rate estimates based on stellar mass are roughly comparable. Assuming a Salpeter-like initial mass function (Salpeter 1955) we find that NGC 4993 has formed $\sim 4 \times 10^{8}$ binaries with stars whose initial masses are greater than 5 $M_{\odot}$. For a typical merger delay time (which dominates the lifetime of a BNS system) of 4 Gyr (see Table 2 in the Appendix) we calculate that NGC 4993's efficiency in forming BNS merger systems per number of massive binaries is in the range of $\sim(1\text{–}50) \times 10^{-4}$.

We model NGC 4993 with a spherically symmetric stellar and dark-matter halo profile. We note that the galaxy type is E/S0, an intermediate morphology between spiral and elliptical galaxies, and it can possibly retain a disk structure component instead of a pure spherical, radial profile (Lambas et al. 1992). Most recently, however, Im et al. (2017) show that NGC 4993 is dominated by its bulge, further supporting our assumption of a spherical gravitational potential. In addition, NGC 4993 has an axis ratio of ≃0.9 (Crook et al. 2007), which is consistent with a nearly spherical elliptical galaxy (though it does not exclude a face-on disk). In our analysis, we also adopt circular orbits for the galactic motion of the progenitors prior to SN2, even though there are more complex orbits allowed in realistic potentials expected for galaxies like NGC 4993 (e.g., box orbits). The key effect of including the galactic orbits is simple: the pre-SN progenitor was already in motion with orbital velocities of hundreds of $\mathrm{km\,s^{-1}}$, which is comparable to the systemic post-SN velocities of the source. The specific shape of the galactic orbits or of the gravitational potential does not appear to be of particular importance, as our constraints are primarily dictated from the necessity that the system remain bound prior to SN2. This assertion is further supported by the fact that our quantitative constraints for progenitor properties are comparable to those found for BNS systems in the Milky Way where the galactic potential of a spiral galaxy is used instead.

In conclusion, we use a minimal set of observational information to constrain GW170817's immediate progenitor, the SN kick imparted to the second neutron star, and its birth location in NGC 4993 with an appropriate galaxy model and the merger offset both informed by photometry. We obtain relatively robust constraints on the progenitor properties, albeit not always tight, strongly influenced by the requirement of keeping the binary bound after the SN and having the merger occur relatively close to the center of NGC 4993. The GW170817 progenitor constraints derived in this study are in good agreement with the progenitor constraints derived for the Galactic BNS systems as well (e.g., Wong et al. 2010; Tauris et al. 2017).

It is important to note that these constraints are essentially unchanged provided the stellar populations in NGC 4993 are older than 1 Gyr. The current literature on NGC 4993 does not provide quantitative information on the galaxy's star formation history. Recent observations (e.g., Foley et al. 2017) might indicate some star formation activity, but as an E/S0 galaxy, it is unlikely that GW170817 was the result of very recent star formation (DeGraaff et al. 2007). Additionally, observations from Im et al. (2017) conclude that the stellar population in NGC 4993 is older than 3 Gyr. Our results strongly indicate that, for a small projected offset like that of GW170817, knowledge of the precise star formation history of the host galaxy is not vital in further constraining SN kicks and progenitor properties.

As more EM counterparts to BNS mergers are identified, we will add to the current sample of BNS systems from the Milky Way and inferred from extragalactic sGRB offset measurements to advance our constraints on progenitor properties. We note that larger projected offsets from the host-galaxy center may provide stronger constraints on the SN-kick magnitudes. In such cases, information on the age of the host-galaxy stellar population, and therefore on the BNS inspiral time, may become more useful.

The authors gratefully acknowledge the support of the United States National Science Foundation (NSF) for the construction and operation of the LIGO Laboratory and Advanced LIGO as well as the Science and Technology Facilities Council (STFC) of the United Kingdom, the Max-Planck-Society (MPS), and the State of Niedersachsen/Germany for support of the construction of Advanced LIGO and construction and operation of the GEO600 detector. Additional support for Advanced LIGO was provided by the Australian Research Council. The authors gratefully acknowledge the Italian Istituto Nazionale di Fisica Nucleare (INFN), the French Centre National de la Recherche Scientifique (CNRS) and the Foundation for Fundamental Research on Matter supported by the Netherlands Organisation for Scientific Research, for the construction and operation of the Virgo detector and the creation and support of the EGO consortium. The authors also gratefully acknowledge research support from these agencies as well as by the Council of Scientific and Industrial Research of India, the Department of Science and Technology, India, the Science & Engineering Research Board (SERB), India, the Ministry of Human Resource Development, India, the Spanish Agencia Estatal de Investigación, the Vicepresidència i Conselleria d'Innovació Recerca i Turisme and the Conselleria d'Educació i Universitat del Govern de les Illes Balears, the Conselleria d'Educació Investigació Cultura i Esport de la Generalitat Valenciana, the National Science Centre of Poland, the Swiss National Science Foundation (SNSF), the Russian Foundation for Basic Research, the Russian Science Foundation, the European Commission, the European Regional Development Funds (ERDF), the Royal Society, the Scottish Funding Council, the Scottish Universities Physics Alliance, the Hungarian Scientific Research Fund (OTKA), the Lyon Institute of Origins (LIO), the National Research, Development and Innovation Office Hungary (NKFI), the National Research Foundation of Korea, Industry Canada and the Province of Ontario through the Ministry of Economic Development and Innovation, the Natural Science and Engineering Research Council Canada, the Canadian Institute for Advanced Research, the Brazilian Ministry of Science, Technology, Innovations, and Communications, the International Center for Theoretical Physics South American Institute for Fundamental Research (ICTP-SAIFR), the Research Grants Council of Hong Kong, the National Natural Science Foundation of China (NSFC), the Leverhulme Trust, the Research Corporation, the Ministry of Science and Technology (*MOST*), Taiwan and the Kavli Foundation. The authors gratefully acknowledge the support of the NSF, STFC, MPS, INFN, CNRS, and the State of Niedersachsen/Germany for provision of computational resources.

## Appendix
## Detailed Constraints and Statistics for Progenitor Properties

We provide more detailed PDFs on progenitor properties inferred from GW170817-like systems with different delay time constraints (Figure 8) and summary statistics for output PDFs (Table 2). In Table 2, we include quantities that measure the degree to which the PDFs change when we constrain the delay times and require that the binaries match the observed offset of GW170817. This PDF comparison is done using the KL divergence (Kullback & Leibler 1951): $\mathrm{KL}(P||Q) = \int p(x)\log[p(x)/q(x)]dx$, where we take $Q$ to be the samples prior to applying a constraint (i.e., delay time or correct offset) and $P$ to be the samples post-application of the constraint.

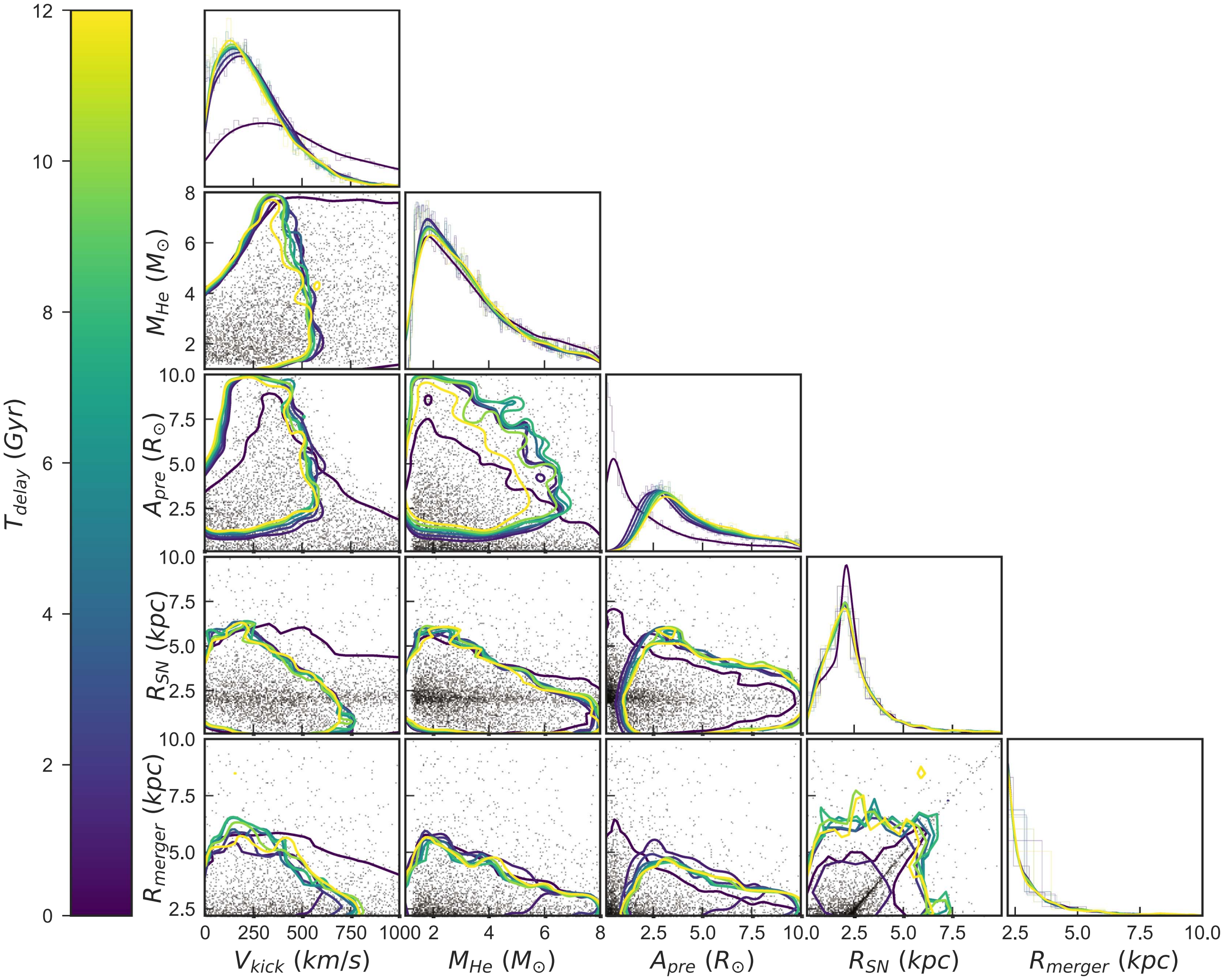

**Figure 8.** Marginalized and joint PDFs on progenitor system properties $V_{\rm kick}$, $M_{\rm He}$, $A_{\rm pre}$, $R_{\rm SN}$, and $R_{\rm merger}$. We restrict GW170817-like systems to various lower limits for $T_{\rm delay}$. The black points show the full population of binaries that correspond to the measured offset of GW170817 (i.e., they have no constraint on $T_{\rm delay}$). As delay times become $\gtrsim$1 Gyr, the constraints on GW170817-like samples are significantly tightened, in particular, removing systems with low $A_{\rm pre}$ and thus short inspiral times, systems with extremely high SN-kick velocities, and systems that are born as BNSs and quickly merge right at the offset of GW170817. The diagonal line in the joint $R_{\rm merger}$–$R_{\rm SN}$ PDF, for example, is an artifact of extremely short inspiral times leading to BNS systems merging at the same location as the second supernova.

**Table 2**
Summary Statistics for Various Input Distributions and Minimum Delay Time Constraints

| Input | | | $T_{\rm delay}$ (Gyr) | | Progenitor Properties | | | | | KL Divergences | | | |
|---|---|---|---|---|---|---|---|---|---|---|---|---|---|
| $V_{\rm kick}$ | $M_{\rm He}$ | $A_{\rm pre}$ | min | max | $V_{\rm kick}$ (km s$^{-1}$) | $M_{\rm He}$ ($M_\odot$) | $A_{\rm pre}$ ($R_\odot$) | $R_{\rm SN}$ (kpc) | $T_{\rm delay}$ (Gyr) | offset | $V_{\rm kick}$, $M_{\rm He}$ | $M_{\rm He}$, $A_{\rm pre}$ | $V_{\rm kick}$, $R_{\rm SN}$ |
| H | U | U | 0 | 14 | $373^{+289}_{-218}$ | $2.83^{+3.71}_{-1.44}$ | $2.64^{+6.00}_{-2.40}$ | $2.08^{+3.74}_{-1.57}$ | $0.14^{+9.39}_{-0.14}$ | 0.19 | ref | ref | ref |
| H | U | U | 1 | 14 | $315^{+240}_{-184}$ | $3.16^{+3.78}_{-1.74}$ | $4.03^{+5.04}_{-2.48}$ | $1.95^{+3.90}_{-1.47}$ | $4.01^{+8.38}_{-2.84}$ | 0.25 | 0.18 | 0.32 | 0.16 |
| H | U | U | 1.8 | 2.2 | $320^{+240}_{-181}$ | $3.06^{+3.74}_{-1.65}$ | $3.65^{+5.36}_{-2.23}$ | $1.94^{+4.04}_{-1.46}$ | $1.99^{+0.19}_{-0.17}$ | 0.49 | 0.69 | 0.83 | 0.59 |
| H | U | L | 0 | 14 | $390^{+328}_{-240}$ | $2.66^{+2.79}_{-1.27}$ | $0.40^{+3.99}_{-0.29}$ | $2.20^{+3.56}_{-1.57}$ | $0.00^{+3.44}_{-0.00}$ | 0.14 | ref | ref | ref |
| H | U | L | 1 | 14 | $305^{+249}_{-184}$ | $3.10^{+3.64}_{-1.68}$ | $2.75^{+5.02}_{-1.53}$ | $1.94^{+3.93}_{-1.48}$ | $3.42^{+8.61}_{-2.29}$ | 0.27 | 0.43 | 1.4 | 0.38 |
| H | U | L | 1.8 | 2.2 | $313^{+240}_{-186}$ | $2.96^{+3.61}_{-1.54}$ | $2.40^{+4.95}_{-1.27}$ | $1.94^{+3.55}_{-1.50}$ | $1.99^{+0.19}_{-0.17}$ | 0.48 | 1.2 | 2.3 | 0.95 |
| H | PL | U | 0 | 14 | $346^{+294}_{-208}$ | $1.75^{+2.00}_{-0.47}$ | $2.26^{+5.95}_{-2.02}$ | $2.20^{+3.79}_{-1.43}$ | $0.08^{+8.63}_{-0.08}$ | 0.16 | ref | ref | ref |
| H | PL | U | 1 | 14 | $285^{+254}_{-173}$ | $1.79^{+2.26}_{-0.51}$ | $3.93^{+4.96}_{-2.37}$ | $2.14^{+3.91}_{-1.39}$ | $3.79^{+8.50}_{-2.64}$ | 0.22 | 0.16 | 0.41 | 0.17 |
| H | PL | U | 1.8 | 2.2 | $296^{+249}_{-175}$ | $1.80^{+2.21}_{-0.52}$ | $3.53^{+5.33}_{-2.12}$ | $2.12^{+3.91}_{-1.41}$ | $1.99^{+0.19}_{-0.17}$ | 0.39 | 0.68 | 0.90 | 0.62 |
| H | PL | L | 0 | 14 | $375^{+321}_{-231}$ | $1.77^{+1.87}_{-0.49}$ | $0.42^{+3.46}_{-0.30}$ | $2.26^{+3.51}_{-1.33}$ | $0.00^{+2.42}_{-0.00}$ | 0.10 | ref | ref | ref |
| H | PL | L | 1 | 14 | $285^{+265}_{-177}$ | $1.81^{+2.23}_{-0.53}$ | $2.73^{+4.87}_{-1.52}$ | $2.14^{+3.93}_{-1.42}$ | $3.25^{+8.63}_{-2.13}$ | 0.22 | 0.32 | 1.5 | 0.41 |
| H | PL | L | 1.8 | 2.2 | $289^{+268}_{-173}$ | $1.82^{+2.14}_{-0.54}$ | $2.38^{+4.90}_{-1.27}$ | $2.14^{+3.97}_{-1.44}$ | $1.98^{+0.20}_{-0.16}$ | 0.37 | 1.1 | 2.4 | 1.1 |
| U | U | U | 0 | 14 | $470^{+1390}_{-420}$ | $2.99^{+4.01}_{-1.59}$ | $1.43^{+6.12}_{-1.27}$ | $2.17^{+3.68}_{-1.60}$ | $0.01^{+8.68}_{-0.01}$ | 0.16 | ref | ref | ref |
| U | U | U | 1 | 14 | $239^{+374}_{-212}$ | $2.79^{+3.71}_{-1.41}$ | $3.40^{+5.35}_{-1.91}$ | $2.05^{+3.88}_{-1.52}$ | $4.22^{+8.27}_{-3.05}$ | 0.26 | 0.47 | 0.66 | 0.47 |
| U | U | U | 1.8 | 2.2 | $260^{+384}_{-222}$ | $2.79^{+3.75}_{-1.40}$ | $2.93^{+5.62}_{-1.58}$ | $2.00^{+3.92}_{-1.53}$ | $1.99^{+0.19}_{-0.17}$ | 0.43 | 0.68 | 1.2 | 0.72 |
| U | U | L | 0 | 14 | $840^{+1390}_{-770}$ | $3.09^{+3.92}_{-1.67}$ | $0.30^{+2.23}_{-0.19}$ | $2.25^{+3.46}_{-1.45}$ | $0.00^{+1.12}_{-0.00}$ | 0.073 | ref | ref | ref |
| U | U | L | 1 | 14 | $211^{+409}_{-190}$ | $2.81^{+3.44}_{-1.41}$ | $2.51^{+4.38}_{-1.30}$ | $2.06^{+3.81}_{-1.53}$ | $3.55^{+8.55}_{-2.42}$ | 0.30 | 1.0 | 2.1 | 1.1 |
| U | U | L | 1.8 | 2.2 | $227^{+378}_{-208}$ | $2.84^{+3.21}_{-1.44}$ | $2.16^{+4.49}_{-1.06}$ | $2.02^{+3.75}_{-1.52}$ | $1.99^{+0.19}_{-0.17}$ | 0.45 | 1.5 | 3.3 | 1.5 |
| U | PL | U | 0 | 14 | $350^{+1230}_{-320}$ | $1.75^{+2.07}_{-0.48}$ | $1.57^{+5.81}_{-1.40}$ | $2.23^{+3.63}_{-1.43}$ | $0.03^{+8.84}_{-0.03}$ | 0.15 | ref | ref | ref |
| U | PL | U | 1 | 14 | $183^{+390}_{-163}$ | $1.76^{+1.94}_{-0.48}$ | $3.45^{+5.06}_{-1.86}$ | $2.22^{+3.74}_{-1.39}$ | $4.03^{+8.42}_{-2.86}$ | 0.20 | 0.33 | 0.64 | 0.35 |
| U | PL | U | 1.8 | 2.2 | $199^{+384}_{-181}$ | $1.75^{+2.00}_{-0.47}$ | $2.88^{+5.58}_{-1.43}$ | $2.23^{+3.46}_{-1.42}$ | $1.99^{+0.19}_{-0.17}$ | 0.35 | 0.58 | 1.1 | 0.60 |
| U | PL | L | 0 | 14 | $630^{+1490}_{-570}$ | $1.79^{+2.15}_{-0.52}$ | $0.31^{+2.42}_{-0.20}$ | $2.28^{+3.44}_{-1.21}$ | $0.00^{+1.41}_{-0.00}$ | 0.075 | ref | ref | ref |
| U | PL | L | 1 | 14 | $175^{+424}_{-159}$ | $1.80^{+1.96}_{-0.53}$ | $2.71^{+4.12}_{-1.42}$ | $2.22^{+3.78}_{-1.42}$ | $3.41^{+8.68}_{-2.28}$ | 0.20 | 0.76 | 2.1 | 0.85 |
| U | PL | L | 1.8 | 2.2 | $189^{+416}_{-171}$ | $1.81^{+1.99}_{-0.54}$ | $2.36^{+3.90}_{-1.15}$ | $2.20^{+3.74}_{-1.46}$ | $1.98^{+0.19}_{-0.17}$ | 0.34 | 1.2 | 3.0 | 1.2 |
| BP | BP | U | 0 | 14 | $7^{+241}_{-4}$ | $1.42^{+1.25}_{-0.17}$ | $2.24^{+2.92}_{-1.94}$ | $2.31^{+3.44}_{-0.87}$ | $0.9^{+10.4}_{-0.9}$ | 0.098 | ref | ref | ref |
| BP | BP | U | 1 | 14 | $7^{+212}_{-4}$ | $1.42^{+1.27}_{-0.17}$ | $3.39^{+2.79}_{-1.35}$ | $2.32^{+3.42}_{-0.81}$ | $4.52^{+8.13}_{-3.33}$ | 0.12 | 0.028 | 0.59 | 0.031 |
| BP | BP | U | 1.8 | 2.2 | $7^{+218}_{-4}$ | $1.42^{+1.25}_{-0.18}$ | $2.76^{+2.46}_{-1.06}$ | $2.31^{+3.41}_{-0.86}$ | $1.99^{+0.19}_{-0.17}$ | 0.21 | 0.28 | 1.9 | 0.26 |
| BP | BP | L | 0 | 14 | $7^{+245}_{-5}$ | $1.43^{+1.35}_{-0.18}$ | $0.59^{+3.11}_{-0.47}$ | $2.31^{+3.39}_{-0.72}$ | $0.01^{+6.24}_{-0.01}$ | 0.073 | ref | ref | ref |
| BP | BP | L | 1 | 14 | $7^{+209}_{-4}$ | $1.42^{+1.35}_{-0.18}$ | $3.07^{+1.59}_{-1.41}$ | $2.32^{+3.45}_{-0.83}$ | $3.66^{+8.55}_{-2.53}$ | 0.11 | 0.064 | 1.6 | 0.079 |
| BP | BP | L | 1.8 | 2.2 | $7^{+216}_{-5}$ | $1.43^{+1.38}_{-0.18}$ | $2.73^{+1.09}_{-1.21}$ | $2.32^{+3.45}_{-0.86}$ | $1.99^{+0.19}_{-0.17}$ | 0.20 | 0.39 | 2.9 | 0.41 |

**Note.** Reported values are the median and 90% confidence interval. The letters in the first three columns indicate the input sampling used: uniform (U), Hobbs (H), log uniform (L), power law (PL), and BP16 (BP); see Section 2.4 for more details. KL divergence scores (in units of Nats) are reported in the four right-most columns, quantifying information gained by imposing constraints (and by proxy, how much the PDFs change). "Offset" quantifies the amount of information learned by taking all binaries that survive the second supernova and imposing the constraint that they merge at the correct projected offset, using the ($V_{\rm kick}$, $M_{\rm He}$) joint PDF. The remaining three KL values take all post-SN binaries with the correct offset and sampling method indicated by the first three columns of a given row and restrict to those with a $T_{\rm delay}$ range specified by that row. These compare the 2D PDFs indicated by the column headers. The rows with $T_{\rm delay} \in [0, 14]$ Gyr have no age constraints, and thus their age-restricted KL divergence will always be zero (they are being compared to themselves). We have labeled these as "ref" instead of zero, to make this clear. As a rough rule of thumb, KL values $\lesssim$0.1 correspond to small differences in the distributions, $\sim$0.4–0.6 to modest differences, and $\gtrsim$1.0 to quite large differences. In general, we find that we learn slightly more by imposing the offset constraint if the age constraint is tighter. We also consistently get higher KL values when $A_{\rm pre}$ is included in the analysis, which is due to its strong correlation with $T_{\rm delay}$. Input distributions are described in detail in Section 2.4.

Specifically, the KL divergence measures the information gained by updating a prior $Q$ to a posterior $P$. The values are computed by histogramming the samples from $Q$ to approximate $q(x)$ and using the same bin locations to make a histogram of $p(x)$. The integral for KL divergence then becomes analytic.

## References

Aasi, J., Abbott, B. P., Abbott, R., et al. 2015, CQGra, 32, 074001
Abadie, J., Abbott, B. P., Abbott, R., et al. 2010, CQGra, 27, 173001
Abbott, B. P., Abbott, R., Abbott, T. D., et al. 2016, PhRvL, 116, 061102
Abbott, B. P., Abbott, R., Abbott, T. D., et al. 2017a, ApJL, 848, L13
Abbott, B. P., Abbott, R., Abbott, T. D., et al. 2017b, PhRvL, 119, 161101
Abbott, B. P., Abbott, R., Abbott, T. D., et al. 2017c, ApJL, 848, L12
Acernese, F., Agathos, M., Agatsuma, K., et al. 2015, CQGra, 32, 024001
Alexander, K., Berger, E., Fong, W., et al. 2017, ApJL, 848, L21
Allam, S., Annis, J., Berger, E., et al. 2017, GRB Coordinates Network, 21530
Arcavi, I., Howell, D. A., McCully, C., et al. 2017a, GCN, 21538
Arcavi, I., Hosseinzadeh, G., Howell, D., et al. 2017b, Natur, 551, 64
Arzoumanian, Z., Chernoff, D. F., & Cordes, J. M. 2002, ApJ, 568, 289
Bae, Y.-B., Kim, C., & Lee, H. M. 2014, MNRAS, 440, 2714
Barr, J. M., Bedregal, A. G., Aragón-Salamanca, A., Merrifield, M. R., & Bamford, S. P. 2007, A&A, 470, 173
Behroozi, P. S., Ramirez-Ruiz, E., & Fryer, C. L. 2014, ApJ, 792, 123
Belczynski, K., Kalogera, V., & Bulik, T. 2002, ApJ, 572, 407
Bellini, A., Bianchini, P., Varri, A. L., et al. 2017, ApJ, 844, 167
Beniamini, P., & Piran, T. 2016, MNRAS, 456, 4089
Berger, E. 2014, ARA&A, 52, 43
Berger, E., Price, P. A., Cenko, S. B., et al. 2005, Natur, 438, 988
Blaauw, A. 1961, BAN, 15, 265
Blanchet, L., Damour, T., Iyer, B. R., Will, C. M., & Wiseman, A. 1995, PhRvL, 74, 3515
Brown, D. A., Harry, I., Lundgren, A., & Nitz, A. H. 2012, PhRvD, 86, 084017

Chatterjee, S., Vlemmings, W. H. T., Brisken, W. F., et al. 2005, ApJL, 630, L61
Church, R. P., Levan, A. J., Davies, M. B., & Tanvir, N. 2011, MNRAS, 413, 2004
Clark, J. P. A., van den Heuvel, E. P. J., & Sutantyo, W. 1979, A&A, 72, 120
Corsi, A., Hallinan, G., Mooley, K., et al. 2017, GCN, 21815
Coulter, D. A., Kilpatrick, C. D., Siebert, M. R., et al. 2017a, GCN, 21529
Coulter, D. A., Kilpatrick, C. D., Siebert, M. R., et al. 2017b, Sci, https://doi.org/10.1126/science.aap9811
Crook, A. C., Huchra, J. P., Martimbeau, N., et al. 2007, ApJ, 655, 790
Cutler, C., & Flanagan, E. E. 1994, PhRvD, 49, 2658
De Loore, C., De Greve, J. P., & de Cuyper, J. P. 1975, Ap&SS, 36, 219
de Vaucouleurs, G. 1948, AnAp, 11, 247
DeGraaff, R. B., Blakeslee, J. P., Meurer, G. R., & Putman, M. E. 2007, ApJ, 671, 1624
Deller, A., Bailes, M., Andreoni, I., et al. 2017a, GCN, 21588
Deller, A., Bailes, M., Andreoni, I., et al. 2017b, GCN, 21850
Deller, A., Bailes, M., Andreoni, I., et al. 2017c, GCN, 21897
Duffy, A. R., Schaye, J., Kay, S. T., & Dalla Vecchia, C. 2008, MNRAS, 390, L64
Eichler, D., Livio, M., Piran, T., & Schramm, D. N. 1989, Natur, 340, 126
Einstein, A. 1916, Sitzungsberichte der Königlich Preußischen Akademie der Wissenschaften (Berlin), 1, 688
Einstein, A. 1918, Sitzungsberichte der Königlich Preußischen Akademie der Wissenschaften (Berlin), 1, 154
Flannery, B. P., & van den Heuvel, E. P. J. 1975, A&A, 39, 61
Foley, R. J., Kilpatrick, C. D., Nicholl, M., & Berger, E. 2017, GCN, 21536
Fong, W., & Berger, E. 2013, ApJ, 776, 18
Fong, W., Berger, E., & Fox, D. B. 2010, ApJ, 708, 9
Fox, D. B., Frail, D. A., Price, P. A., et al. 2005, Natur, 437, 845
Freedman, W. L., Madore, B. F., Gibson, B. K., et al. 2001, ApJ, 553, 47
Gehrels, N., Sarazin, C. L., O'Brien, P. T., et al. 2005, Natur, 437, 851
Goldstein, A., Veres, P., Burns, E., et al. 2017, ApJL, 848, L14
Goodman, J. 1986, ApJL, 308, L47
Gott, J. R., III, Gunn, J. E., & Ostriker, J. P. 1970, ApJL, 160, L91
Grindlay, J., Portegies Zwart, S., & McMillan, S. 2006, NatPh, 2, 116
Haggard, D., Nynka, M., Kalogera, V., et al. 2017c, Sci, https://doi.org/10.1126/science.aap9855
Haggard, D., Nynka, M., Kalogera, V., Evans, P., & Cenko, B. 2017b, GCN, 21798
Haggard, D., Nynka, M., Ruan, J. J., et al. 2017a, ApJL, 848, L25
Harris, W. E. 2016, AJ, 151, 102
Harris, W. E., Harris, G. L., & Hudson, M. J. 2015, ApJ, 806, 36
Hernquist, L. 1990, ApJ, 356, 359
Hjorth, J., Watson, D., Fynbo, J. P. U., et al. 2005, Natur, 437, 859
Hobbs, G., Lorimer, D. R., Lyne, A. G., & Kramer, M. 2005, MNRAS, 360, 974
Huchra, J. P., Macri, L. M., Masters, K. L., et al. 2012, ApJS, 199, 26
Hulse, R. A., & Taylor, J. H. 1975, ApJL, 195, L51
Im, M., Yoon, Y., Lee, S. K. J., et al. 2017, ApJL, 849, L16
Ivanova, N., Belczynski, K., Kalogera, V., Rasio, F. A., & Taam, R. E. 2003, ApJ, 592, 475
Ivanova, N., Heinke, C. O., Rasio, F. A., Belczynski, K., & Fregeau, J. M. 2008, MNRAS, 386, 553
Janka, H.-T. 2013, MNRAS, 434, 1355
Janka, H.-T., Langanke, K., Marek, A., Martínez-Pinedo, G., & Müller, B. 2007, PhR, 442, 38
Kalogera, V. 1996, ApJ, 471, 352
Kalogera, V., Belczynski, K., Kim, C., O'Shaughnessy, R., & Willems, B. 2007, PhR, 442, 75
Kalogera, V., & Lorimer, D. R. 2000, ApJ, 530, 890
Kalogera, V., & Webbink, R. F. 1998, ApJ, 493, 351
Kasliwal, M., Nakar, E., Singer, L. P., & Kaplan, D. E. A. 2017, Sci, https://doi.org/10.1126/science.aap9455
Kaspi, V. M., Bailes, M., Manchester, R. N., Stappers, B. W., & Bell, J. F. 1996, Natur, 381, 584
Kouveliotou, C., Meegan, C. A., Fishman, G. J., et al. 1993, ApJL, 413, L101
Kullback, S., & Leibler, R. A. 1951, Ann. Math. Stat., 22, 79
Kusenko, A., & Segrè, G. 1996, PhRvL, 77, 4872
Lambas, D. G., Maddox, S. J., & Loveday, J. 1992, MNRAS, 258, 404
Lauberts, A. 1982, ESO/Uppsala survey of the ESO(B) Atlas (Munich: ESO)
Lauberts, A., & Valentijn, E. A. 1989, The Surface Photometry Catalogue of the ESO-Uppsala Galaxies (Munich: ESO)
Licquia, T. C., & Newman, J. A. 2015, ApJ, 806, 96
LIGO Scientific Collaboration & Virgo Collaboration 2017, GCN, 21513
Lim, S. H., Mo, H. J., Lu, Y., Wang, H., & Yang, X. 2017, MNRAS, 470, 2982
Lipunov, V. M., Gorbovskoy, E., Kornilov, V. G., et al. 2017b, GCN, 21570
Lipunov, V., Gorbovskoy, E., Kornilov, V., et al. 2017a, ApJL, 850, L1
Lyne, A. G., & Lorimer, D. R. 1994, Natur, 369, 127
Massevitch, A. G., Tutukov, A. V., & Iungelson, L. R. 1976, Ap&SS, 40, 115
Mooley, K. P., Hallinan, G., Corsi, A., et al. 2017, GCN, 21814
Narayan, R., Paczynski, B., & Piran, T. 1992, ApJL, 395, L83
Navarro, J. F., Frenk, C. S., & White, S. D. M. 1996, ApJ, 462, 563
Osłowski, S., Bulik, T., Gondek-Rosińska, D., & Belczyński, K. 2011, MNRAS, 413, 461
Paczynski, B. 1986, ApJL, 308, L43
Peters, P. C. 1964, PhRv, 136, 1224
Piran, T., & Shaviv, N. J. 2005, PhRvL, 94, 051102
Planck Collaboration, Ade, P. A. R., Aghanim, N., et al. 2016, A&A, 594, A13
Podsiadlowski, P., Langer, N., Poelarends, A. J. T., et al. 2004, ApJ, 612, 1044
Pols, O. R., & Dewi, J. D. M. 2002, PASA, 19, 233
Postnov, K. A., & Yungelson, L. R. 2014, LRR, 17, 3
Salpeter, E. E. 1955, ApJ, 121, 161
Savchenko, V., Ferrigno, C., Kuulkers, E., et al. 2017, ApJL, 848, L23
Schaye, J., Crain, R. A., Bower, R. G., et al. 2015, MNRAS, 446, 521
Soares-Santos, M., Holz, D., Annis, J., et al. 2017, ApJL, 848, L16
Stairs, I. H., Thorsett, S. E., Dewey, R. J., Kramer, M., & McPhee, C. A. 2006, MNRAS, 373, L50
Tanvir, N. R., & Levan, A. J. 2017, GCN, 21544
Tauris, T. M., Kramer, M., Freire, P. C. C., et al. 2017, ApJ, 846, 170
Taylor, J. H., & Weisberg, J. M. 1982, ApJ, 253, 908
Troja, E., King, A. R., O'Brien, P. T., Lyons, N., & Cusumano, G. 2008, MNRAS, 385, L10
Tutukov, A., & Yungelson, L. 1973, NInfo, 27, 86
van den Heuvel, E. P. J. 2007, in AIP Conf. Ser. 924, The Multicolored Landscape of Compact Objects and Their Explosive Origins, ed. T. di Salvo et al. (Melville, NY: AIP), 598
van den Heuvel, E. P. J., & De Loore, C. 1973, A&A, 25, 387
van den Heuvel, E. P. J., & Heise, J. 1972, NPhS, 239, 67
Verbunt, F., Igoshev, A., & Cator, E. 2017, arXiv:1708.08281
Villasenor, J. S., Lamb, D. Q., Ricker, G. R., et al. 2005, Natur, 437 855
Willems, B., & Kalogera, V. 2004, ApJL, 603, L101
Willems, B., Kaplan, J., Fragos, T., Kalogera, V., & Belczynski, K. 2006, PhRvD, 74, 043003
Wong, T.-W., Willems, B., & Kalogera, V. 2010, ApJ, 721, 1689
Yang, S., Valenti, S., Sand, D., et al. 2017, GCN, 21531
Yu, P. C., Ngeow, C. C., Ip, W. H., et al. 2017, GCN, 21669

B. P. Abbott[1], R. Abbott[1], T. D. Abbott[2], F. Acernese[3,4], K. Ackley[5,6], C. Adams[7], T. Adams[8], P. Addesso[9], R. X. Adhikari[1], V. B. Adya[10], C. Affeldt[10], M. Afrough[11], B. Agarwal[12], M. Agathos[13], K. Agatsuma[14], N. Aggarwal[15], O. D. Aguiar[16], L. Aiello[17,18], A. Ain[19], P. Ajith[20], B. Allen[10,21,22], G. Allen[12], A. Allocca[23,24], P. A. Altin[25], A. Amato[26], A. Ananyeva[1], S. B. Anderson[1], W. G. Anderson[21], S. V. Angelova[27], S. Antier[28], S. Appert[1], K. Arai[1], M. C. Araya[1], J. S. Areeda[29], N. Arnaud[28,30], K. G. Arun[31], S. Ascenzi[32,33], G. Ashton[10], M. Ast[34], S. M. Aston[7], P. Astone[35], D. V. Atallah[36], P. Aufmuth[22], C. Aulbert[10], K. AultONeal[37], C. Austin[2], A. Avila-Alvarez[29], S. Babak[38], P. Bacon[39], M. K. M. Bader[14], S. Bae[40], P. T. Baker[41], F. Baldaccini[42,43], G. Ballardin[30], S. W. Ballmer[44], S. Banagiri[45], J. C. Barayoga[1], S. E. Barclay[46], B. C. Barish[1], D. Barker[47], K. Barkett[48], F. Barone[3,4], B. Barr[46], L. Barsotti[15], M. Barsuglia[39], D. Barta[49], J. Bartlett[47], I. Bartos[5,50], R. Bassiri[51], A. Basti[23,24],

J. C. Batch[47], M. Bawaj[43,52], J. C. Bayley[46], M. Bazzan[53,54], B. Bécsy[55], C. Beer[10], M. Bejger[56], I. Belahcene[28], A. S. Bell[46], B. K. Berger[1], G. Bergmann[10], J. J. Bero[57], C. P. L. Berry[58], D. Bersanetti[59], A. Bertolini[14], J. Betzwieser[7], S. Bhagwat[44], R. Bhandare[60], I. A. Bilenko[61], G. Billingsley[1], C. R. Billman[5], J. Birch[7], R. Birney[62], O. Birnholtz[10], S. Biscans[1,15], S. Biscoveanu[6,63], A. Bisht[22], M. Bitossi[24,30], C. Biwer[44], M. A. Bizouard[28], J. K. Blackburn[1], J. Blackman[48], C. D. Blair[1,64], D. G. Blair[64], R. M. Blair[47], S. Bloemen[65], O. Bock[10], N. Bode[10], M. Boer[66], G. Bogaert[66], A. Bohe[38], F. Bondu[67], E. Bonilla[51], R. Bonnand[8], B. A. Boom[14], R. Bork[1], V. Boschi[24,30], S. Bose[19,68], K. Bossie[7], Y. Bouffanais[39], A. Bozzi[30], C. Bradaschia[24], P. R. Brady[21], M. Branchesi[17,18], J. E. Brau[69], T. Briant[70], A. Brillet[66], M. Brinkmann[10], V. Brisson[28], P. Brockill[21], J. E. Broida[71], A. F. Brooks[1], D. D. Brown[72], S. Brunett[1], C. C. Buchanan[2], A. Buikema[15], T. Bulik[73], H. J. Bulten[14,74], A. Buonanno[38,75], D. Buskulic[8], C. Buy[39], R. L. Byer[51], M. Cabero[10], L. Cadonati[76], G. Cagnoli[26,77], C. Cahillane[1], J. Calderón Bustillo[76], T. A. Callister[1], E. Calloni[4,78], J. B. Camp[79], M. Canepa[59,80], P. Canizares[65], K. C. Cannon[81], H. Cao[72], J. Cao[82], C. D. Capano[10], E. Capocasa[39], F. Carbognani[30], S. Caride[83], M. F. Carney[84], J. Casanueva Diaz[28], C. Casentini[32,33], S. Caudill[14,21], M. Cavaglià[11], F. Cavalier[28], R. Cavalieri[30], G. Cella[24], C. B. Cepeda[1], P. Cerdá-Durán[85], G. Cerretani[23,24], E. Cesarini[33,86], S. J. Chamberlin[63], M. Chan[46], S. Chao[87], P. Charlton[88], E. Chase[89], E. Chassande-Mottin[39], D. Chatterjee[21], B. D. Cheeseboro[41], H. Y. Chen[90], X. Chen[64], Y. Chen[48], H.-P. Cheng[5], H. Chia[5], A. Chincarini[59], A. Chiummo[30], T. Chmiel[84], H. S. Cho[91], M. Cho[75], J. H. Chow[25], N. Christensen[66,71], Q. Chu[64], A. J. K. Chua[13], S. Chua[70], A. K. W. Chung[92], S. Chung[64], G. Ciani[5,53,54], R. Ciolfi[93,94], C. E. Cirelli[51], A. Cirone[59,80], F. Clara[47], J. A. Clark[76], P. Clearwater[95], F. Cleva[66], C. Cocchieri[11], E. Coccia[17,18], P.-F. Cohadon[70], D. Cohen[28], A. Colla[35,96], C. G. Collette[97], L. R. Cominsky[98], M. Constancio, Jr.[16], L. Conti[54], S. J. Cooper[58], P. Corban[7], T. R. Corbitt[2], I. Cordero-Carrión[99], K. R. Corley[50], A. Corsi[83], S. Cortese[30], C. A. Costa[16], M. W. Coughlin[1,71], S. B. Coughlin[89], J.-P. Coulon[66], S. T. Countryman[50], P. Couvares[1], P. B. Covas[100], E. E. Cowan[76], D. M. Coward[64], M. J. Cowart[7], D. C. Coyne[1], R. Coyne[83], J. D. E. Creighton[21], T. D. Creighton[101], J. Cripe[2], S. G. Crowder[102], T. J. Cullen[2,29], A. Cumming[46], L. Cunningham[46], E. Cuoco[30], T. Dal Canton[79], G. Dálya[55], S. L. Danilishin[10,22], S. D'Antonio[33], K. Danzmann[10,22], A. Dasgupta[103], C. F. Da Silva Costa[5], V. Dattilo[30], I. Dave[60], M. Davier[28], D. Davis[44], E. J. Daw[104], B. Day[76], S. De[44], D. DeBra[51], J. Degallaix[26], M. De Laurentis[4,17], S. Deléglise[70], W. Del Pozzo[23,24,58], N. Demos[15], T. Denker[10], T. Dent[10], R. De Pietri[105,106], V. Dergachev[38], R. De Rosa[4,78], R. T. DeRosa[7], C. De Rossi[26,30], R. DeSalvo[107], O. de Varona[10], J. Devenson[27], S. Dhurandhar[19], M. C. Díaz[101], L. Di Fiore[4], M. Di Giovanni[94,108], T. Di Girolamo[4,50,78], A. Di Lieto[23,24], S. Di Pace[35,96], I. Di Palma[35,96], F. Di Renzo[23,24], Z. Doctor[90], V. Dolique[26], F. Donovan[15], K. L. Dooley[11], S. Doravari[10], I. Dorrington[36], R. Douglas[46], M. Dovale Álvarez[58], T. P. Downes[21], M. Drago[10], C. Dreissigacker[10], J. C. Driggers[47], Z. Du[82], M. Ducrot[8], P. Dupej[46], S. E. Dwyer[47], T. B. Edo[104], M. C. Edwards[71], A. Effler[7], H.-B. Eggenstein[10,38], P. Ehrens[1], J. Eichholz[1], S. S. Eikenberry[5], R. A. Eisenstein[15], R. C. Essick[15], D. Estevez[8], Z. B. Etienne[41], T. Etzel[1], M. Evans[15], T. M. Evans[7], M. Factourovich[50], V. Fafone[17,32,33], H. Fair[44], S. Fairhurst[36], X. Fan[82], S. Farinon[59], B. Farr[90], W. M. Farr[58], E. J. Fauchon-Jones[36], M. Favata[109], M. Fays[36], C. Fee[84], H. Fehrmann[10], J. Feicht[1], M. M. Fejer[51], A. Fernandez-Galiana[15], I. Ferrante[23,24], E. C. Ferreira[16], F. Ferrini[30], F. Fidecaro[23,24], D. Finstad[44], I. Fiori[30], D. Fiorucci[39], M. Fishbach[90], R. P. Fisher[44], M. Fitz-Axen[45], R. Flaminio[26,110], M. Fletcher[46], H. Fong[111], J. A. Font[85,112], P. W. F. Forsyth[25], S. S. Forsyth[76], J.-D. Fournier[66], S. Frasca[35,96], F. Frasconi[24], Z. Frei[55], A. Freise[58], R. Frey[69], V. Frey[28], E. M. Fries[1], P. Fritschel[15], V. V. Frolov[7], P. Fulda[5], M. Fyffe[7], H. Gabbard[46], B. U. Gadre[19], S. M. Gaebel[58], J. R. Gair[113], L. Gammaitoni[42], M. R. Ganija[72], S. G. Gaonkar[19], C. Garcia-Quiros[100], F. Garufi[4,78], B. Gateley[47], S. Gaudio[37], G. Gaur[114], V. Gayathri[115], N. Gehrels[79,162], G. Gemme[59], E. Genin[30], A. Gennai[24], D. George[12], J. George[60], L. Gergely[116], V. Germain[8], S. Ghonge[76], Abhirup Ghosh[20], Archisman Ghosh[14,20], S. Ghosh[14,21,65], J. A. Giaime[2,7], K. D. Giardina[7], A. Giazotto[24], K. Gill[37], L. Glover[107], E. Goetz[117], R. Goetz[5], S. Gomes[36], B. Goncharov[6], J. M. Gonzalez Castro[23,24], A. Gopakumar[118], M. L. Gorodetsky[61], S. E. Gossan[1], M. Gosselin[30], R. Gouaty[8], A. Grado[4,119], C. Graef[46], M. Granata[26], A. Grant[46], S. Gras[15], C. Gray[47], G. Greco[120,121], A. C. Green[58], E. M. Gretarsson[37], P. Groot[65], H. Grote[10], S. Grunewald[38], P. Gruning[28], G. M. Guidi[120,121], X. Guo[82], A. Gupta[63], M. K. Gupta[103], K. E. Gushwa[1], E. K. Gustafson[1], R. Gustafson[117], O. Halim[17,18], B. R. Hall[68], E. D. Hall[15], E. Z. Hamilton[36], G. Hammond[46], M. Haney[122], M. M. Hanke[10], J. Hanks[47], C. Hanna[63], M. D. Hannam[36], O. A. Hannuksela[92], J. Hanson[7], T. Hardwick[2], J. Harms[17,18], G. M. Harry[123], I. W. Harry[38], M. J. Hart[46], C.-J. Haster[111], K. Haughian[46], J. Healy[57], A. Heidmann[70], M. C. Heintze[7], H. Heitmann[66], P. Hello[28], G. Hemming[30], M. Hendry[46], I. S. Heng[46], J. Hennig[46], A. W. Heptonstall[1], M. Heurs[10,22], S. Hild[46], T. Hinderer[65], D. Hoak[30], D. Hofman[26], A. M. Holgado[12], K. Holt[7], D. E. Holz[90], P. Hopkins[36], C. Horst[21], J. Hough[46], E. A. Houston[46], E. J. Howell[64], A. Hreibi[66], Y. M. Hu[10], E. A. Huerta[12], D. Huet[28], B. Hughey[37], S. Husa[100], S. H. Huttner[46], T. Huynh-Dinh[7], N. Indik[10], R. Inta[83], G. Intini[35,96], H. N. Isa[46], J.-M. Isac[70], M. Isi[1], B. R. Iyer[20], K. Izumi[47], T. Jacqmin[70], K. Jani[76], P. Jaranowski[124], S. Jawahar[62], F. Jiménez-Forteza[100], W. W. Johnson[2], D. I. Jones[125], R. Jones[46], R. J. G. Jonker[14], L. Ju[64], J. Junker[10], C. V. Kalaghatgi[36], V. Kalogera[89], B. Kamai[1], S. Kandhasamy[7], G. Kang[40], J. B. Kanner[1], S. J. Kapadia[21], S. Karki[69], K. S. Karvinen[10], M. Kasprzack[2], M. Katolik[12], E. Katsavounidis[15], W. Katzman[7], S. Kaufer[22], K. Kawabe[47], F. Kéfélian[66], D. Keitel[46], A. J. Kemball[12], R. Kennedy[104], C. Kent[36], J. S. Key[126], F. Y. Khalili[61], I. Khan[17,33], S. Khan[10], Z. Khan[103], E. A. Khazanov[127], N. Kijbunchoo[25], Chunglee Kim[128], J. C. Kim[129], K. Kim[92], W. Kim[72], W. S. Kim[130], Y.-M. Kim[91], C. Kimball[89], S. J. Kimbrell[76], E. J. King[72], P. J. King[47], M. Kinley-Hanlon[123], R. Kirchhoff[10], J. S. Kissel[47], L. Kleybolte[34], S. Klimenko[5], T. D. Knowles[41], P. Koch[10], S. M. Koehlenbeck[10], S. Koley[14], V. Kondrashov[1],

A. Kontos[15], M. Korobko[34], W. Z. Korth[1], I. Kowalska[73], D. B. Kozak[1], C. Krämer[10], V. Kringel[10], A. Królak[131,132], G. Kuehn[10], P. Kumar[111], R. Kumar[103], S. Kumar[20], L. Kuo[87], A. Kutynia[131], S. Kwang[21], B. D. Lackey[38], K. H. Lai[92], M. Landry[47], R. N. Lang[133], J. Lange[57], B. Lantz[51], R. K. Lanza[15], S. L. Larson[89], A. Lartaux-Vollard[28], P. D. Lasky[6], M. Laxen[7], A. Lazzarini[1], C. Lazzaro[54], P. Leaci[35,96], S. Leavey[46], C. H. Lee[91], H. K. Lee[134], H. M. Lee[135], H. W. Lee[129], K. Lee[46], J. Lehmann[10], A. Lenon[41], M. Leonardi[94,108], N. Leroy[28], N. Letendre[8], Y. Levin[6], T. G. F. Li[92], S. D. Linker[107], T. B. Littenberg[136], J. Liu[64], R. K. L. Lo[92], N. A. Lockerbie[62], L. T. London[36], J. E. Lord[44], M. Lorenzini[17,18], V. Loriette[137], M. Lormand[7], G. Losurdo[24], J. D. Lough[10], C. O. Lousto[57], G. Lovelace[29], H. Lück[10,22], D. Lumaca[32,33], A. P. Lundgren[10], R. Lynch[15], Y. Ma[48], R. Macas[36], S. Macfoy[27], B. Machenschalk[10], M. MacInnis[15], D. M. Macleod[36], I. Magaña Hernandez[21], F. Magaña-Sandoval[44], L. Magaña Zertuche[44], R. M. Magee[63], E. Majorana[35], I. Maksimovic[137], N. Man[66], V. Mandic[45], V. Mangano[46], G. L. Mansell[25], M. Manske[21,25], M. Mantovani[30], F. Marchesoni[43,52], F. Marion[8], S. Márka[50], Z. Márka[50], C. Markakis[12], A. S. Markosyan[51], A. Markowitz[1], E. Maros[1], A. Marquina[99], F. Martelli[120,121], L. Martellini[66], I. W. Martin[46], R. M. Martin[109], D. V. Martynov[15], K. Mason[15], E. Massera[104], A. Masserot[8], T. J. Massinger[1], M. Masso-Reid[46], S. Mastrogiovanni[35,96], A. Matas[45], F. Matichard[1,15], L. Matone[50], N. Mavalvala[15], N. Mazumder[68], R. McCarthy[47], D. E. McClelland[25], S. McCormick[7], L. McCuller[15], S. C. McGuire[138], G. McIntyre[1], J. McIver[1], D. J. McManus[25], L. McNeill[6], T. McRae[25], S. T. McWilliams[41], D. Meacher[63], G. D. Meadors[10,38], M. Mehmet[10], J. Meidam[14], E. Mejuto-Villa[9], A. Melatos[95], G. Mendell[47], R. A. Mercer[21], E. L. Merilh[47], M. Merzougui[66], S. Meshkov[1], C. Messenger[46], C. Messick[63], R. Metzdorff[70], P. M. Meyers[45], H. Miao[58], C. Michel[26], H. Middleton[58], E. E. Mikhailov[139], L. Milano[4,78], A. L. Miller[5,35,96], B. B. Miller[89], J. Miller[15], M. Millhouse[140], M. C. Milovich-Goff[107], O. Minazzoli[66,141], Y. Minenkov[33], J. Ming[38], C. Mishra[142], S. Mitra[19], V. P. Mitrofanov[61], G. Mitselmakher[5], R. Mittleman[15], D. Moffa[84], A. Moggi[24], K. Mogushi[11], M. Mohan[30], S. R. P. Mohapatra[15], M. Montani[120,121], C. J. Moore[13], D. Moraru[47], G. Moreno[47], S. R. Morriss[101], B. Mours[8], C. M. Mow-Lowry[58], G. Mueller[5], A. W. Muir[36], Arunava Mukherjee[10], D. Mukherjee[21], S. Mukherjee[101], N. Mukund[19], A. Mullavey[7], J. Munch[72], E. A. Muñiz[44], M. Muratore[37], P. G. Murray[46], K. Napier[76], I. Nardecchia[32,33], L. Naticchioni[35,96], R. K. Nayak[143], J. Neilson[107], G. Nelemans[14,65], T. J. N. Nelson[7], M. Nery[10], A. Neunzert[117], L. Nevin[1], J. M. Newport[123], G. Newton[46,163], K. K. Y. Ng[92], T. T. Nguyen[25], D. Nichols[65], A. B. Nielsen[10], S. Nissanke[14,65], A. Nitz[10], A. Noack[10], F. Nocera[30], D. Nolting[7], C. North[36], L. K. Nuttall[36], J. Oberling[47], G. D. O'Dea[107], G. H. Ogin[144], J. J. Oh[130], S. H. Oh[130], F. Ohme[10], M. A. Okada[16], M. Oliver[100], P. Oppermann[10], Richard J. Oram[7], B. O'Reilly[7], R. Ormiston[45], L. F. Ortega[5], R. O'Shaughnessy[57], S. Ossokine[38], D. J. Ottaway[72], H. Overmier[7], B. J. Owen[83], A. E. Pace[63], J. Page[136], M. A. Page[64], A. Pai[115,145], S. A. Pai[60], J. R. Palamos[69], O. Palashov[127], C. Palomba[35], A. Pal-Singh[34], Howard Pan[87], Huang-Wei Pan[87], B. Pang[48], P. T. H. Pang[92], C. Pankow[89], F. Pannarale[36], B. C. Pant[60], F. Paoletti[24], A. Paoli[30], M. A. Papa[10,21,38], A. Parida[19], W. Parker[7], D. Pascucci[46], A. Pasqualetti[30], R. Passaquieti[23,24], D. Passuello[24], M. Patil[132], B. Patricelli[24,146], B. L. Pearlstone[46], M. Pedraza[1], R. Pedurand[26,147], L. Pekowsky[44], A. Pele[7], S. Penn[148], C. J. Perez[47], A. Perreca[1,94,108], L. M. Perri[89], H. P. Pfeiffer[38,111], M. Phelps[46], O. J. Piccinni[35,96], M. Pichot[66], F. Piergiovanni[120,121], V. Pierro[9], G. Pillant[30], L. Pinard[26], I. M. Pinto[9], M. Pirello[47], M. Pitkin[46], M. Poe[21], R. Poggiani[23,24], P. Popolizio[30], E. K. Porter[39], A. Post[10], J. Powell[46,149], J. Prasad[19], J. W. W. Pratt[37], G. Pratten[100], V. Predoi[36], T. Prestegard[21], M. Prijatelj[10], M. Principe[9], S. Privitera[38], G. A. Prodi[94,108], L. G. Prokhorov[61], O. Puncken[10], M. Punturo[43], P. Puppo[35], M. Pürrer[38], H. Qi[21], V. Quetschke[101], E. A. Quintero[1], R. Quitzow-James[69], D. S. Rabeling[25], H. Radkins[47], P. Raffai[55], S. Raja[60], C. Rajan[60], B. Rajbhandari[83], M. Rakhmanov[101], K. E. Ramirez[101], A. Ramos-Buades[100], P. Rapagnani[35,96], V. Raymond[38], M. Razzano[23,24], J. Read[29], T. Regimbau[66], L. Rei[59], S. Reid[62], D. H. Reitze[1,5], W. Ren[12], S. D. Reyes[44], F. Ricci[35,96], P. M. Ricker[12], S. Rieger[10], K. Riles[117], M. Rizzo[57], N. A. Robertson[1,46], R. Robie[46], F. Robinet[28], A. Rocchi[33], L. Rolland[8], J. G. Rollins[1], V. J. Roma[69], R. Romano[3,4], C. L. Romel[47], J. H. Romie[7], D. Rosińska[56,150], M. P. Ross[151], S. Rowan[46], A. Rüdiger[10], P. Ruggi[30], G. Rutins[27], K. Ryan[47], S. Sachdev[1], T. Sadecki[47], L. Sadeghian[21], M. Sakellariadou[152], L. Salconi[30], M. Saleem[115], F. Salemi[10], A. Samajdar[143], L. Sammut[6], L. M. Sampson[89], E. J. Sanchez[1], L. E. Sanchez[1], N. Sanchis-Gual[85], V. Sandberg[47], J. R. Sanders[44], B. Sassolas[26], B. S. Sathyaprakash[36,63], O. Sauter[117], R. L. Savage[47], A. Sawadsky[34], P. Schale[69], M. Scheel[48], J. Scheuer[89], J. Schmidt[10], P. Schmidt[1,65], R. Schnabel[34], R. M. S. Schofield[69], A. Schönbeck[34], E. Schreiber[10], D. Schuette[10,22], B. W. Schulte[10], B. F. Schutz[10,36], S. G. Schwalbe[37], J. Scott[46], S. M. Scott[25], E. Seidel[12], D. Sellers[7], A. S. Sengupta[153], D. Sentenac[30], V. Sequino[17,32,33], A. Sergeev[127], D. A. Shaddock[25], T. J. Shaffer[47], A. A. Shah[136], M. S. Shahriar[89], M. B. Shaner[107], L. Shao[38], B. Shapiro[51], P. Shawhan[75], A. Sheperd[21], D. H. Shoemaker[15], D. M. Shoemaker[76], K. Siellez[76], X. Siemens[21], M. Sieniawska[56], D. Sigg[47], A. D. Silva[16], L. P. Singer[79], A. Singh[10,22,38], A. Singhal[17,35], A. M. Sintes[100], B. J. J. Slagmolen[25], B. Smith[7], J. R. Smith[29], R. J. E. Smith[1,6], S. Somala[154], E. J. Son[130], J. A. Sonnenberg[21], B. Sorazu[46], F. Sorrentino[59], T. Souradeep[19], A. P. Spencer[46], A. K. Srivastava[103], K. Staats[37], A. Staley[50], M. Steinke[10], J. Steinlechner[34,46], S. Steinlechner[34], D. Steinmeyer[10], S. P. Stevenson[58,149], R. Stone[101], D. J. Stops[58], K. A. Strain[46], G. Stratta[120,121], S. E. Strigin[61], A. Strunk[47], R. Sturani[155], A. L. Stuver[7], T. Z. Summerscales[156], L. Sun[95], S. Sunil[103], J. Suresh[19], P. J. Sutton[36], B. L. Swinkels[30], M. J. Szczepańczyk[37], M. Tacca[14], S. C. Tait[46], C. Talbot[6], D. Talukder[69], D. B. Tanner[5], M. Tápai[116], A. Taracchini[38], J. D. Tasson[71], J. A. Taylor[136], R. Taylor[1], S. V. Tewari[148], T. Theeg[10], F. Thies[10], E. G. Thomas[58], M. Thomas[7], P. Thomas[47], K. A. Thorne[7], E. Thrane[6], S. Tiwari[17,94], V. Tiwari[36], K. V. Tokmakov[62], K. Toland[46], M. Tonelli[23,24], Z. Tornasi[46], A. Torres-Forné[85], C. I. Torrie[1], D. Töyrä[58], F. Travasso[30,43], G. Traylor[7], J. Trinastic[5], M. C. Tringali[94,108],

L. Trozzo[24,157], K. W. Tsang[14], M. Tse[15], R. Tso[1], L. Tsukada[81], D. Tsuna[81], D. Tuyenbayev[101], K. Ueno[21], D. Ugolini[158], C. S. Unnikrishnan[118], A. L. Urban[1], S. A. Usman[36], H. Vahlbruch[22], G. Vajente[1], G. Valdes[2], N. van Bakel[14], M. van Beuzekom[14], J. F. J. van den Brand[14,74], C. Van Den Broeck[14,159], D. C. Vander-Hyde[44], L. van der Schaaf[14], J. V. van Heijningen[14], A. A. van Veggel[46], M. Vardaro[53,54], V. Varma[48], S. Vass[1], M. Vasúth[49], A. Vecchio[58], G. Vedovato[54], J. Veitch[46], P. J. Veitch[72], K. Venkateswara[151], G. Venugopalan[1], D. Verkindt[8], F. Vetrano[120,121], A. Viceré[120,121], A. D. Viets[21], S. Vinciguerra[58], D. J. Vine[27], J.-Y. Vinet[66], S. Vitale[15], T. Vo[44], H. Vocca[42,43], C. Vorvick[47], S. P. Vyatchanin[61], A. R. Wade[1], L. E. Wade[84], M. Wade[84], R. Walet[14], M. Walker[29], L. Wallace[1], S. Walsh[10,21,38], G. Wang[17,121], H. Wang[58], J. Z. Wang[63], W. H. Wang[101], Y. F. Wang[92], R. L. Ward[25], J. Warner[47], M. Was[8], J. Watchi[97], B. Weaver[47], L.-W. Wei[10,22], M. Weinert[10], A. J. Weinstein[1], R. Weiss[15], L. Wen[64], E. K. Wessel[12], P. Weßels[10], J. Westerweck[10], T. Westphal[10], K. Wette[25], J. T. Whelan[57], B. F. Whiting[5], C. Whittle[6], D. Wilken[10], D. Williams[46], R. D. Williams[1], A. R. Williamson[65], J. L. Willis[1,160], B. Willke[10,22], M. H. Wimmer[10], W. Winkler[10], C. C. Wipf[1], H. Wittel[10,22], G. Woan[46], J. Woehler[10], J. Wofford[57], K. W. K. Wong[92], J. Worden[47], J. L. Wright[46], D. S. Wu[10], D. M. Wysocki[57], S. Xiao[1], H. Yamamoto[1], C. C. Yancey[75], L. Yang[161], M. J. Yap[25], M. Yazback[5], Hang Yu[15], Haocun Yu[15], M. Yvert[8], A. Zadrożny[131], M. Zanolin[37], T. Zelenova[30], J.-P. Zendri[54], M. Zevin[89], L. Zhang[1], M. Zhang[139], T. Zhang[46], Y.-H. Zhang[57], C. Zhao[64], M. Zhou[89], Z. Zhou[89], S. J. Zhu[10,38], X. J. Zhu[6], M. E. Zucker[1,15], and J. Zweizig[1]
(LIGO Scientific Collaboration and Virgo Collaboration)

[1] LIGO, California Institute of Technology, Pasadena, CA 91125, USA
[2] Louisiana State University, Baton Rouge, LA 70803, USA
[3] Università di Salerno, Fisciano, I-84084 Salerno, Italy
[4] INFN, Sezione di Napoli, Complesso Universitario di Monte S.Angelo, I-80126 Napoli, Italy
[5] University of Florida, Gainesville, FL 32611, USA
[6] OzGrav, School of Physics & Astronomy, Monash University, Clayton, VIC 3800, Australia
[7] LIGO Livingston Observatory, Livingston, LA 70754, USA
[8] Laboratoire d'Annecy-le-Vieux de Physique des Particules (LAPP), Université Savoie Mont Blanc, CNRS/IN2P3, F-74941 Annecy, France
[9] University of Sannio at Benevento, I-82100 Benevento, Italy and INFN, Sezione di Napoli, I-80100 Napoli, Italy
[10] Max Planck Institute for Gravitational Physics (Albert Einstein Institute), D-30167 Hannover, Germany
[11] The University of Mississippi, University, MS 38677, USA
[12] NCSA, University of Illinois at Urbana-Champaign, Urbana, IL 61801, USA
[13] University of Cambridge, Cambridge CB2 1TN, UK
[14] Nikhef, Science Park, 1098 XG Amsterdam, The Netherlands
[15] LIGO, Massachusetts Institute of Technology, Cambridge, MA 02139, USA
[16] Instituto Nacional de Pesquisas Espaciais, 12227-010 São José dos Campos, São Paulo, Brazil
[17] Gran Sasso Science Institute (GSSI), I-67100 L'Aquila, Italy
[18] INFN, Laboratori Nazionali del Gran Sasso, I-67100 Assergi, Italy
[19] Inter-University Centre for Astronomy and Astrophysics, Pune 411007, India
[20] International Centre for Theoretical Sciences, Tata Institute of Fundamental Research, Bengaluru 560089, India
[21] University of Wisconsin–Milwaukee, Milwaukee, WI 53201, USA
[22] Leibniz Universität Hannover, D-30167 Hannover, Germany
[23] Università di Pisa, I-56127 Pisa, Italy
[24] INFN, Sezione di Pisa, I-56127 Pisa, Italy
[25] OzGrav, Australian National University, Canberra, ACT 0200, Australia
[26] Laboratoire des Matériaux Avancés (LMA), CNRS/IN2P3, F-69622 Villeurbanne, France
[27] SUPA, University of the West of Scotland, Paisley PA1 2BE, UK
[28] LAL, Univ. Paris-Sud, CNRS/IN2P3, Université Paris-Saclay, F-91898 Orsay, France
[29] California State University Fullerton, Fullerton, CA 92831, USA
[30] European Gravitational Observatory (EGO), I-56021 Cascina, Pisa, Italy
[31] Chennai Mathematical Institute, Chennai 603103, India
[32] Università di Roma Tor Vergata, I-00133 Roma, Italy
[33] INFN, Sezione di Roma Tor Vergata, I-00133 Roma, Italy
[34] Universität Hamburg, D-22761 Hamburg, Germany
[35] INFN, Sezione di Roma, I-00185 Roma, Italy
[36] Cardiff University, Cardiff CF24 3AA, UK
[37] Embry-Riddle Aeronautical University, Prescott, AZ 86301, USA
[38] Max Planck Institute for Gravitational Physics (Albert Einstein Institute), D-14476 Potsdam-Golm, Germany
[39] APC, AstroParticule et Cosmologie, Université Paris Diderot, CNRS/IN2P3, CEA/Irfu, Observatoire de Paris, Sorbonne Paris Cité, F-75205 Paris Cedex 13, France
[40] Korea Institute of Science and Technology Information, Daejeon 34141, Korea
[41] West Virginia University, Morgantown, WV 26506, USA
[42] Università di Perugia, I-06123 Perugia, Italy
[43] INFN, Sezione di Perugia, I-06123 Perugia, Italy
[44] Syracuse University, Syracuse, NY 13244, USA
[45] University of Minnesota, Minneapolis, MN 55455, USA
[46] SUPA, University of Glasgow, Glasgow G12 8QQ, UK
[47] LIGO Hanford Observatory, Richland, WA 99352, USA
[48] Caltech CaRT, Pasadena, CA 91125, USA
[49] Wigner RCP, RMKI, H-1121 Budapest, Konkoly Thege Miklós út 29-33, Hungary
[50] Columbia University, New York, NY 10027, USA
[51] Stanford University, Stanford, CA 94305, USA
[52] Università di Camerino, Dipartimento di Fisica, I-62032 Camerino, Italy

[53] Università di Padova, Dipartimento di Fisica e Astronomia, I-35131 Padova, Italy
[54] INFN, Sezione di Padova, I-35131 Padova, Italy
[55] Institute of Physics, Eötvös University, Pázmány P. s. 1/A, Budapest 1117, Hungary
[56] Nicolaus Copernicus Astronomical Center, Polish Academy of Sciences, 00-716, Warsaw, Poland
[57] Rochester Institute of Technology, Rochester, NY 14623, USA
[58] University of Birmingham, Birmingham B15 2TT, UK
[59] INFN, Sezione di Genova, I-16146 Genova, Italy
[60] RRCAT, Indore MP 452013, India
[61] Faculty of Physics, Lomonosov Moscow State University, Moscow 119991, Russia
[62] SUPA, University of Strathclyde, Glasgow G1 1XQ, UK
[63] The Pennsylvania State University, University Park, PA 16802, USA
[64] OzGrav, University of Western Australia, Crawley, WA 6009, Australia
[65] Department of Astrophysics/IMAPP, Radboud University Nijmegen, P.O. Box 9010, 6500 GL Nijmegen, The Netherlands
[66] Artemis, Université Côte d'Azur, Observatoire Côte d'Azur, CNRS, CS 34229, F-06304 Nice Cedex 4, France
[67] Institut FOTON, CNRS, Université de Rennes 1, F-35042 Rennes, France
[68] Washington State University, Pullman, WA 99164, USA
[69] University of Oregon, Eugene, OR 97403, USA
[70] Laboratoire Kastler Brossel, UPMC-Sorbonne Universités, CNRS, ENS-PSL Research University, Collège de France, F-75005 Paris, France
[71] Carleton College, Northfield, MN 55057, USA
[72] OzGrav, University of Adelaide, Adelaide, SA 5005, Australia
[73] Astronomical Observatory Warsaw University, 00-478 Warsaw, Poland
[74] VU University Amsterdam, 1081 HV Amsterdam, The Netherlands
[75] University of Maryland, College Park, MD 20742, USA
[76] Center for Relativistic Astrophysics, Georgia Institute of Technology, Atlanta, GA 30332, USA
[77] Université Claude Bernard Lyon 1, F-69622 Villeurbanne, France
[78] Università di Napoli "Federico II," Complesso Universitario di Monte S.Angelo, I-80126 Napoli, Italy
[79] NASA Goddard Space Flight Center, Greenbelt, MD 20771, USA
[80] Dipartimento di Fisica, Università degli Studi di Genova, I-16146 Genova, Italy
[81] RESCEU, University of Tokyo, Tokyo 113-0033, Japan
[82] Tsinghua University, Beijing 100084, China
[83] Texas Tech University, Lubbock, TX 79409, USA
[84] Kenyon College, Gambier, OH 43022, USA
[85] Departamento de Astronomía y Astrofísica, Universitat de València, E-46100 Burjassot, València, Spain
[86] Museo Storico della Fisica e Centro Studi e Ricerche Enrico Fermi, I-00184 Roma, Italy
[87] National Tsing Hua University, Hsinchu City, 30013 Taiwan, Republic of China
[88] Charles Sturt University, Wagga Wagga, NSW 2678, Australia
[89] Center for Interdisciplinary Exploration & Research in Astrophysics (CIERA), Northwestern University, Evanston, IL 60208, USA
[90] University of Chicago, Chicago, IL 60637, USA
[91] Pusan National University, Busan 46241, Korea
[92] The Chinese University of Hong Kong, Shatin, NT, Hong Kong
[93] INAF, Osservatorio Astronomico di Padova, I-35122 Padova, Italy
[94] INFN, Trento Institute for Fundamental Physics and Applications, I-38123 Povo, Trento, Italy
[95] OzGrav, University of Melbourne, Parkville, VIC 3010, Australia
[96] Università di Roma "La Sapienza," I-00185 Roma, Italy
[97] Université Libre de Bruxelles, B-1050 Brussels, Belgium
[98] Sonoma State University, Rohnert Park, CA 94928, USA
[99] Departamento de Matemáticas, Universitat de València, E-46100 Burjassot, València, Spain
[100] Universitat de les Illes Balears, IAC3—IEEC, E-07122 Palma de Mallorca, Spain
[101] The University of Texas Rio Grande Valley, Brownsville, TX 78520, USA
[102] Bellevue College, Bellevue, WA 98007, USA
[103] Institute for Plasma Research, Bhat, Gandhinagar 382428, India
[104] The University of Sheffield, Sheffield S10 2TN, UK
[105] Dipartimento di Scienze Matematiche, Fisiche e Informatiche, Università di Parma, I-43124 Parma, Italy
[106] INFN, Sezione di Milano Bicocca, Gruppo Collegato di Parma, I-43124 Parma, Italy
[107] California State University, Los Angeles, 5151 State University Drive, Los Angeles, CA 90032, USA
[108] Università di Trento, Dipartimento di Fisica, I-38123 Povo, Trento, Italy
[109] Montclair State University, Montclair, NJ 07043, USA
[110] National Astronomical Observatory of Japan, 2-21-1 Osawa, Mitaka, Tokyo 181-8588, Japan
[111] Canadian Institute for Theoretical Astrophysics, University of Toronto, Toronto, ON M5S 3H8, Canada
[112] Observatori Astronòmic, Universitat de València, E-46980 Paterna, València, Spain
[113] School of Mathematics, University of Edinburgh, Edinburgh EH9 3FD, UK
[114] University and Institute of Advanced Research, Koba Institutional Area, Gandhinagar Gujarat 382007, India
[115] IISER-TVM, CET Campus, Trivandrum Kerala 695016, India
[116] University of Szeged, Dóm tér 9, Szeged 6720, Hungary
[117] University of Michigan, Ann Arbor, MI 48109, USA
[118] Tata Institute of Fundamental Research, Mumbai 400005, India
[119] INAF, Osservatorio Astronomico di Capodimonte, I-80131, Napoli, Italy
[120] Università degli Studi di Urbino "Carlo Bo," I-61029 Urbino, Italy
[121] INFN, Sezione di Firenze, I-50019 Sesto Fiorentino, Firenze, Italy
[122] Physik-Institut, University of Zurich, Winterthurerstrasse 190, 8057 Zurich, Switzerland
[123] American University, Washington, DC 20016, USA
[124] University of Białystok, 15-424 Białystok, Poland
[125] University of Southampton, Southampton SO17 1BJ, UK
[126] University of Washington Bothell, 18115 Campus Way NE, Bothell, WA 98011, USA
[127] Institute of Applied Physics, Nizhny Novgorod, 603950, Russia

[128] Korea Astronomy and Space Science Institute, Daejeon 34055, Korea
[129] Inje University Gimhae, South Gyeongsang 50834, Korea
[130] National Institute for Mathematical Sciences, Daejeon 34047, Korea
[131] NCBJ, 05-400 Świerk-Otwock, Poland
[132] Institute of Mathematics, Polish Academy of Sciences, 00656 Warsaw, Poland
[133] Hillsdale College, Hillsdale, MI 49242, USA
[134] Hanyang University, Seoul 04763, Korea
[135] Seoul National University, Seoul 08826, Korea
[136] NASA Marshall Space Flight Center, Huntsville, AL 35811, USA
[137] ESPCI, CNRS, F-75005 Paris, France
[138] Southern University and A&M College, Baton Rouge, LA 70813, USA
[139] College of William and Mary, Williamsburg, VA 23187, USA
[140] Montana State University, Bozeman, MT 59717, USA
[141] Centre Scientifique de Monaco, 8 quai Antoine Ier, MC-98000, Monaco
[142] Indian Institute of Technology Madras, Chennai 600036, India
[143] IISER-Kolkata, Mohanpur, West Bengal 741252, India
[144] Whitman College, 345 Boyer Avenue, Walla Walla, WA 99362 USA
[145] Indian Institute of Technology Bombay, Powai, Mumbai, Maharashtra 400076, India
[146] Scuola Normale Superiore, Piazza dei Cavalieri 7, I-56126 Pisa, Italy
[147] Université de Lyon, F-69361 Lyon, France
[148] Hobart and William Smith Colleges, Geneva, NY 14456, USA
[149] OzGrav, Swinburne University of Technology, Hawthorn VIC 3122, Australia
[150] Janusz Gil Institute of Astronomy, University of Zielona Góra, 65-265 Zielona Góra, Poland
[151] University of Washington, Seattle, WA 98195, USA
[152] King's College London, University of London, London WC2R 2LS, UK
[153] Indian Institute of Technology, Gandhinagar Ahmedabad Gujarat 382424, India
[154] Indian Institute of Technology Hyderabad, Sangareddy, Khandi, Telangana 502285, India
[155] International Institute of Physics, Universidade Federal do Rio Grande do Norte, Natal RN 59078-970, Brazil
[156] Andrews University, Berrien Springs, MI 49104, USA
[157] Università di Siena, I-53100 Siena, Italy
[158] Trinity University, San Antonio, TX 78212, USA
[159] Van Swinderen Institute for Particle Physics and Gravity, University of Groningen, Nijenborgh 4, 9747 AG Groningen, The Netherlands
[160] Abilene Christian University, Abilene, TX 79699, USA
[161] Colorado State University, Fort Collins, CO 80523, USA
[162] Deceased, 2017 February.
[163] Deceased, 2016 December.